\documentclass[fleqn,usenatbib]{mnras}

\usepackage{newtxtext,newtxmath}
\usepackage[T1]{fontenc}

\DeclareRobustCommand{\VAN}[3]{#2}
\let\VANthebibliography\thebibliography
\def\thebibliography{\DeclareRobustCommand{\VAN}[3]{##3}\VANthebibliography}

\usepackage{graphicx}	% Including figure files
\usepackage{amsmath}	% Advanced maths commands

\title[PAH Deficit in a Distant Starburst Core]{A PAH deficit in the starburst core of a distant spiral galaxy}

\author[Z. Liu et al.]{
Zhaoxuan Liu,$^{1,2,3,4}$\thanks{E-mail: zhaoxuan.liu@ipmu.jp}
John D. Silverman,$^{1,2,3,5}$
Emanuele Daddi,$^{4}$
Boris S. Kalita,$^{1,3,6}$\thanks{Joint-Kavli Astrophysics Fellow}
Annagrazia Puglisi,$^{7}$\thanks{Anniversary Fellow}
\newauthor
Qinyue Fei,$^{1,6,8}$
Alvio Renzini,$^{9}$
Daichi Kashino,$^{10}$
Francesco Valentino,$^{11,12}$
Jeyhan S. Kartaltepe,$^{13}$
\newauthor
Daizhong Liu,$^{14}$
Pablo G. P\'erez-Gonz\'alez,$^{15}$
Jed McKinney,$^{16}$\thanks{NASA Hubble Fellow}
Caitlin M. Casey,$^{17,11}$
Xuheng Ding,$^{17}$
\newauthor
Andreas Faisst,$^{18}$
Maximilien Franco,$^{16}$
Darshan Kakkad,$^{19}$
Anton M. Koekemoer,$^{20}$
Erini Lambrides,$^{21}$
\newauthor
Steven Gillman,$^{11,22}$
Ghassem Gozaliasl,$^{23,24}$
Henry Joy McCracken,$^{25}$
Jason Rhodes,$^{26}$
Brant E. Robertson,$^{27}$
\newauthor
Giulia Rodighiero,$^{9,28}$
Wiphu Rujopakarn,$^{29,30}$
Tomoko L. Suzuki,$^{1,3}$
Takumi S. Tanaka,$^{1,2,3}$
\newauthor
Brittany N. Vanderhoof,$^{20}$
Aswin P. Vijayan,$^{31}$
Olivia R. Cooper,$^{16}$\thanks{NSF Graduate Research Fellow}
Aidan Kaminsky,$^{32}$
\newauthor
Georgios E. Magdis$^{11,22,12}$ and
Namrata Roy$^{5}$
\\\\
Affiliations are listed at the end of the paper
}

\date{Accepted XXX. Received YYY; in original form ZZZ}

\pubyear{\the\year{}}

\begin{document}
\label{firstpage}
\pagerange{\pageref{firstpage}--\pageref{lastpage}}
\maketitle

\begin{abstract}
We present high-resolution and spatially-matched observations with JWST and ALMA of a starburst galaxy (PACS-830) at $z=1.46$. The NIRCam observations mainly trace the stellar light while the CO ($J$=5--4) observations map the dense molecular gas at kpc scales. Both datasets reveal the morphology to be that of a gas/dust rich bulge with two extending arms, together resembling a grand-design spiral galaxy. The more pronounced arm contributes 21 $\pm$ 6\% of the total CO emission. 
These results demonstrate that starburst activity at high redshift can be triggered, without undergoing a highly disruptive major merger. We assess the strength and distribution of star formation using two tracers: (1) Polycyclic Aromatic Hydrocarbons (PAHs) emission detected at $8~\mu$m ($L_8$) with a MIRI/F1800W image, and (2) $L_\mathrm{IR}$, inferred from the CO ($J$=5--4) map. The spatial profiles of the $L_\mathrm{IR}$ and $L_8$ are dissimilar, thus leading to a significant deficit of mid-IR ($L_8$) emission in the nucleus. We hypothesize that this is due to the destruction of PAH molecules by the intense ionizing radiation field or decreased emission in the photodissociation region, as seen in nearby star-forming regions and consistent with the galaxy-wide properties of distant starbursts. This study reveals spatial variations in the $L_8$ to $L_\mathrm{IR}$ ratio for the first time at $z>1$, in agreement with expectations from theory. Our analysis underscores the pivotal role of joint high-resolution observations with JWST and ALMA in discerning the different phases of the interstellar medium (ISM) and revealing internal physics in galaxy substructures.
\end{abstract}


%%%%%%%%%%%%%%%%%%%%%%%%%%%%%%%%%%%%%%%%%%%%%%%%%%

%%%%%%%%%%%%%%%%% BODY OF PAPER %%%%%%%%%%%%%%%%%%

\section{Introduction} \label{sec:intro}
Cosmic noon ($1<z<3$), the epoch during which global star formation peaks across cosmic time \citep{madau_cosmic_2014,tacconi_evolution_2020}, is crucial for star-forming galaxies (SFGs) as they develop their internal structure, particularly the bulge \citep[e.g.,][]{tan_situ_2024}. A key distinction between SFGs at this epoch and those in the local Universe is their abundant gas content \citep{tacconi_phibss_2013}, which likely leads to an increased star formation efficiency (SFE), i.e., the star formation rate (SFR) per unit gas mass.

Within the star-forming population even at high-z, the starbursts are rapidly building the bulk of their stellar mass and standing out as significant outliers ($>0.3$ dex) with elevated SFRs compared to typical 'Main Sequence' (MS) galaxies, which is defined by the tight correlation between SFR and stellar mass \citep{daddi_multiwavelength_2007} observed up to $z \sim 10$ \citep{calabro_evolution_2024}. Such intense star formation is associated with dust enrichment and subsequent attenuation of UV/optical light thus requiring tracers of star-formation that can penetrate the heavily obscured regions of galaxies.

At submillimeter wavelengths, starburst galaxies exhibit strong emission lines (e.g., CO and C) and dust continuum, which effectively traces not only star formation activity, less hampered by dust attenuation, but also gas kinematics. Evidently, ALMA is effective at unveiling the nature of starbursts; for instance, kpc-scale studies reveal that many high-z starbursts are rotation-supported disks \citep{hodge_kiloparsec-scale_2016,cowie_submillimeter_2018,gullberg_alma_2019,hodge_high-redshift_2020,rizzo_alma-alpaka_2023}, contrasting with their local analogs, which are predominantly triggered by highly disruptive major mergers. Even for extreme starbursts like hyper-luminous infrared galaxies ($L_\mathrm{IR}>10^{13}\,L_{\sun}$), recent hundred-parsec ALMA studies have found a massive rotating disk at $z=2.3$ instead of major merger origin \citep{liu_detailed_2024}. This growing number of rotationally supported starbursts suggests that secular evolution with in-situ disk instabilities may be sufficient to drive high levels of star formation \citep[see][for a review]{forster_schreiber_star-forming_2020}. However, before the advent of JWST, the limited infrared (IR) capabilities of the last generation of telescopes, e.g., HST, did not allow for spatially-resolved analysis of the stellar counterpart for these rotating disks due to the strong attenuation by dust.

Now with JWST, the full scale of stellar morphology at $z>1$ is being revealed, complete with spiral arms, clumps, and bulges as seen by high resolution and deep NIRCam observations \citep[e.g.,][]{rujopakarn_jwst_2023,liu_jwst_2024,kalita_rest-frame_2023,kalita_near-ir_2024,kalita_clumps_2024,mckinney_scubadive_2024}. In fact, massive galaxies, previously classified as mergers, are now being found to be more typical and disk-like with spiral arms and clumps surrounding bright, compact sub-mm sources emitted from their dust-obscured centers, likely bulges in formation \citep{hodge_kiloparsec-scale_2016,hodge_alma_2019,tan_situ_2024,liu_jwst_2024,polletta_jwsts_2024}.

Another diagnostic opportunity now offered by JWST concerns the resolved imaging and spectroscopy covering Polycyclic aromatic hydrocarbons (PAHs) emissions \citep[e.g.,][]{alberts_smiles_2024,shivaei_new_2024,shivaei_tight_2024}. PAHs are key constituents of interstellar dust in galaxies \citep{draine_infrared_2007}. The emission lines of PAHs, spanning the rest-frame infrared spectrum of 3-20 $\mu$m, serve as efficient coolants and dominate the mid-IR luminosity for high SFR galaxies with little AGN contribution \citep[][and references
therein]{draine_interstellar_2003,tielens_interstellar_2008,li_spitzers_2020}.
Though the attenuation can be non-zero across the mid-IR, PAH features, especially the dominating $7.7\mu m $ line, are less affected by dust obscuration, compared to the traditional H$\alpha$ tracer for dusty, distant SFGs \citep{lai_all_2020,lai_spectroscopic_2024}. Consequently, PAH emissions are utilized to trace SFRs in normal, star-forming galaxies due to their prevalence in star-forming regions illuminated by ultraviolet (UV) photons from young stars \citep[e.g.,][]{calzetti_star_2013,dale_phangs-jwst_2023,chastenet_phangs-jwst_2023,lee_phangs-jwst_2023,ronayne_ceers_2024}. 

While PAH features exhibit a tight correlation with galaxy-wide SFR, whether they are reliable tracers of spatially-resolved star formation is still under investigation. For example, PAH molecules can be dissociated by UV photons from star formation or AGN accretion disk as shown by theory \citep{leger_photo-thermo-dissociation_1989}. Using the Spitzer Space Telescope, many studies have reported a deficit of PAH emission in the cores of active galactic nuclei (AGNs) or compact starbursts based on infrared spectrograph (IRS) spectroscopy \citep[e.g. ][]{desai_pah_2007,diaz-santos_spatial_2011,stierwalt_mid-infrared_2013,stierwalt_mid-infrared_2014,zhang_evidence_2022}. \citet{inami_mid-infrared_2013} also present similar observations in a survey of 202 local luminous infrared galaxies via the AKARI telescope. Recently, studies using mid-infrared instrument imaging (MIRI) aboard JWST have found evidence of PAH deficits in HII regions in nearby AGNs \citep[e.g.,][]{lai_goals-jwst_2022,lai_goals-jwst_2023} and even in normal SFGs \citep[e.g.,][]{egorov_phangsjwst_2023,spilker_spatial_2023, pedrini_feedback_2024, battisti_constraining_2025}. Therefore, such deficits could also occur due to strong radiation fields in star-forming regions and their effects should be considered when utilizing PAH emissions to infer SFRs.

In the distant Universe, such deficits are linked to a high star formation rate density ($\Sigma_\mathrm{SFR}$) and starburstiness ($R_\mathrm{SB}=\mathrm{SFR/SFR_{MS}}$), shown by an empirical relationship established between the global total infrared luminosity $L_\mathrm{IR}$ and PAH emission at rest-frame 8~$\mu m$ $L_{8}$ up to $z\sim2.5$ \citep{elbaz_goodsherschel_2011, elbaz_starbursts_2018}. Such trends are further confirmed by spectroscopic studies using Spitzer \citep[e.g.,][]{pope_probing_2013,mckinney_measuring_2020}. With JWST, MIRI can now provide unparalleled sensitivity and spatial resolution at mid-infrared wavelengths, making resolved studies of galaxies in the distant Universe feasible \citep{magnelli_ceers_2023,ronayne_ceers_2024,alberts_smiles_2024,lyu_primer_2024,shivaei_new_2024}.

Here, we report a spatially-resolved study of a starburst galaxy at $z = 1.463$, PACS-830, with clear spiral morphology and potential bulge in formation, to compare star-forming conditions across the surface of the galaxy. In Section~\ref{sec:data}, we introduce our multi-wavelength observations with HST, JWST, and ALMA. We integrate these datasets to model the dust-obscured stellar mass distribution on a pixel-by-pixel basis using HST and JWST/NIRCam images and to compare the distribution of two SFR tracers, the PAH emissions from JWST/MIRI observations with ALMA CO ($J=$5--4) data. In Section~\ref{sec:modeling}, we detail this modeling process. In Section~\ref{sec:results}, we analyze the results in terms of galaxy morphology and the variation of $L_\mathrm{IR}/L_8$. In this work, we adopt a \citet{chabrier_galactic_2003} IMF and the standard $\Lambda\mathrm{CDM}$ model with $H_{0}=70 \mathrm{~km} \mathrm{~s}^{-1} \mathrm{Mpc}^{-1}, \Omega_{\Lambda}=0.7, \Omega_{\mathrm{M}}=0.3$.

\begin{figure*}
\centering
\includegraphics[width=14cm]{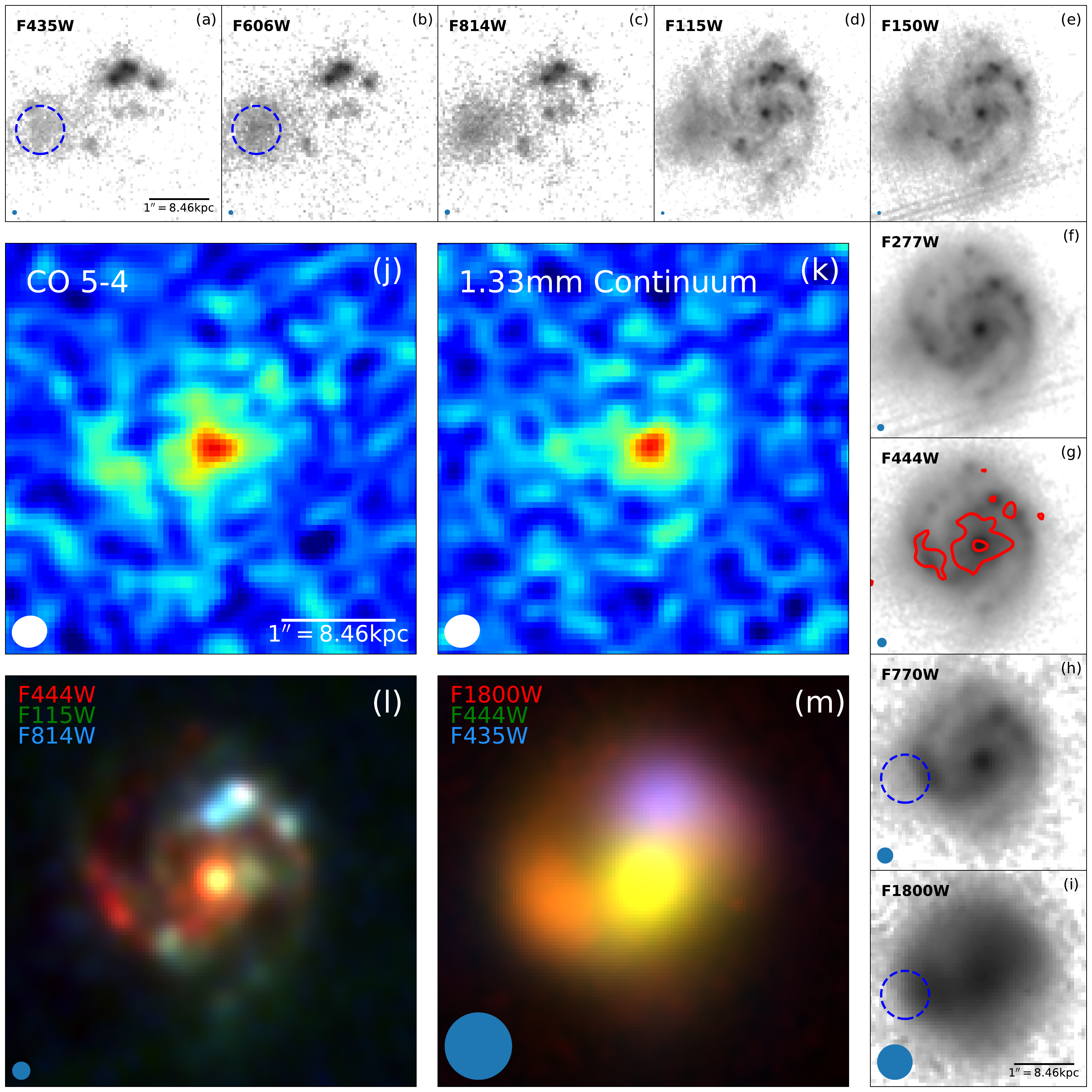}
\caption{Observations of PACS-830. The surrounding panels show images of different instrument/filter combinations: (a) HST F435W, (b) HST F606W, (c) HST F814W, (d) JWST/NIRCam F115W, (e) F150W, (f) F277W, (g) F444W, and (h) MIRI F770W and (i) F1800W images. ALMA observations of PACS-830 are shown in the larger upper panels (j: CO 5--4, k: 1.33mm continuum). All panels span the same region of the sky, with north at the top and east to the left. The PSF of each filter is shown in the lower-left corner of each panel. The physical scale of 1 arcsec is 8.46 kpc, as indicated in the lower-right corner in panels a, i, and j. The bottom two panels in the middle are two RGB images generated by the filter pairs: F444W, F115W and F814W (rest-frame NIR+optical+UV,l) and F1800W, F444W, F435W (PAH+stellar mass+unobscured SFR, m) respectively, after subtracting the foreground galaxy and being convolved to the widest point spread function (PSF) within the pair (see Section~\ref{sec:p2p}). The position of the foreground galaxy is highlighted with blue circles in the panels a-b and h-i. We highlight the emergence of clumpy spiral features at longer wavelengths in the JWST images, which aligns with the CO ($J$=5--4) contours shown in panel g. These red contours are plotted at levels of 3 and 9$\,\times\,\sigma_{\mathrm{rms}}$, where $\sigma_{\mathrm{rms}} = 0.029\,\mathrm{Jy}\,\mathrm{beam}^{-1}\,\mathrm{km\,s}^{-1}$.}
\label{fig:observations}
\end{figure*}

\section{Observations of PACS-830} \label{sec:data}

PACS-830 (J2000 RA = 10:00:08.7462, DEC = +02:19:01.876) is a starburst galaxy, identified by Herschel/PACS photometry \citep{rodighiero_lesser_2011}. SED fitting, based on photometry from UV to radio, including measurements from Herschel, places this galaxy as a moderate starburst, $4\times$ above the MS at z=1.463 (Figure 1 in \citealt{silverman_higher_2015}). With accurate redshift estimation through spectroscopic observations with Subaru Telescope as part of the FMOS-COSMOS survey \citep{Kashino2019}, the total SFR in PACS-830 is estimated to be 412$^{+62}_{-54}$  M$_{\odot}$ yr$^{-1}$, derived from its infrared luminosity $L_\mathrm{IR}$ \citep{silverman_molecular_2018}. Stellar mass ($M_{*}$) estimation using global multi-wavelength photometry by \citet{liu_co_2021} finds  $M_*=10^{11.04}~M_\odot$.

PACS-830 shows no evidence of an AGN. The BPT diagram analysis by \citet{silverman_higher_2015} with FMOS spectra confirms that PACS-830 lies within the star-forming region. The galaxy shows no detection in previous X-ray surveys with $Chandra$ and XMM-$Newton$ in the COSMOS field \citep{cappelluti_xmm-newton_2009, civano_chandra_2016, marchesi_chandra_2016}. Further, \citet{liu_co_2021} conducted SED fitting with 11 bands from Mid-IR to radio using both \texttt{CIGALE} \citep{boquien_cigale_2019} and \texttt{MICHI2} \citep{liu_automated_2019} and found an insignificant AGN contribution.

\subsection{High-resolution ALMA observations}\label{sec:alma_data}

PACS-830 was observed with ALMA Band 6 during a Cycle 4 program (Proposal \#2016.1.01426.S; PI J. Silverman), to map the molecular gas distribution in three starbursts at $z\sim1.5$. The on-source times with extended (47 antennas) and compact (43 antennas) configurations were 70.6 and 11.9 minutes, respectively. The native spectral resolution is 5.022~km/s.

We recalibrated the data by running the Common Astronomy Software Applications (CASA) standard pipelines and co-add the data from the extended and compact configurations to create a data cube with high dynamic range. Before constructing the images, we first measured the centroid and FWHM (Full Width at Half Maximum) of CO (J=5-4) line from the high-resolution data cube provided as a part of the QA2 products from ALMA. After flagging a spectral region of 4$\times$FWHM (305~km/s) centered at 233.96~GHz (line centroid), the continuum was imaged with the \textit{tclean} algorithm in multifrequency synthesis (\textit{mfs}, \citealt{1990MNRAS.246..490C}) mode using the rest channels from 215~GHz to 235~GHz (Figure \ref{fig:observations}.k). We then subtracted the continuum from the visibilities using \textit{uvcontsub}. For the CO ($J$=5--4) intensity map (Figure \ref{fig:observations}.j), we adopted the method outlined in \citet{akins_alma_2022}, i.e.,  running \textit{tclean} in the \textit{mfs} mode over a frequency range covering the emission line to derive a pseudo-moment 0 map. The frequency range of the combined channels is 2$\times$FWHM centered at the line centroid. This approach increased the signal-to-noise ratio (S/N) and ensured the detection of the faint and diffuse emission of the ISM in PACS-830 as it stacks the channels prior to \textit{tclean}. For both images, we utilized \textit{auto-multithresh} to create masks for cleaning with the threshold parameters for extended configurations suggested in \citet{kepley_auto-multithresh_2020}, together with a \textit{multiscale} \citep{cornwell_multiscale_2008} deconvolver to recover the extended emissions. We opted for robust Briggs weighting \citep[0.5;][]{1995AAS...18711202B} to balance the S/N and spatial resolution. We first created the dirty images without cleaning to measure the $\sigma_{\mathrm{rms}}$ and then cleaned the images to $0.5\times\sigma_{\mathrm{rms}}$. The resulting restored beams are 0$^{\prime \prime}$.31 $\times$ 0$^{\prime \prime}$.28 and 0$^{\prime \prime}$.31 $\times$ 0$^{\prime \prime}$.27. We use CASA 4.7.2, as specified in the QA2 report, to carry out the recalibration of the raw data. The imaging, measurements, and analysis, are processed by CASA 6.2.1.

\subsection{JWST and HST observations}

With PACS-830 located in the COSMOS field, it is covered by two JWST Cycle 1 programs: COSMOS-Web (PIs: Casey \& Kartaltepe; \citealt{casey_cosmos-web_2023}; GO \#1727) and PRIMER (PI: Dunlop; \citealt{2021jwst.prop.1837D}; GO \#1837). PACS-830 was observed with four NIRCam filters (F115W, F150W, F277W, F444W; 5$\sigma$ depths: 27.13, 27.35, 27.99, 27.83 magnitudes) and one MIRI filter (F770W; 5$\sigma$ depth: 25.4 magnitudes) by COSMOS-Web and additional MIRI filters (F770W and F1800W; 5$\sigma$ depths: 26.0 and 23.0 magnitudes) by PRIMER. Since the detector angles differ between the two programs, we are unable to stack the F770W observations. Consequently, we use the PRIMER dataset for MIRI imaging observations, as it offers more depth compared to the COSMOS-Web survey. The NIRCam images were recalibrated and produced by the COSMOS-Web team (Franco in prep.) and the data reduction of the
MIRI images from the PRIMER team is detailed in \citet{perez-gonzalez_what_2024}. The NIRCam images (Figure~\ref{fig:observations}.d-g) are all reduced to a spatial scale of 30 mas/pixel while the MIRI ones (Figure~\ref{fig:observations}.h-i) are in 60 mas/pixel. 

We also include rest-frame UV imaging from archival HST/ACS observations in filters F435W, F606W, and F814W, where much of this data comes from the original COSMOS HST \citep{koekemoer_cosmos_2007}, and the CANDELS survey \citep{grogin_candels_2011,koekemoer_candels_2011}, in addition to other archival programs. These images (Figure~\ref{fig:observations}.a-c) were retrieved from the DAWN JWST Archive (DJA) \footnote{https://dawn-cph.github.io/dja/index.html} and reduced to 40 mas/pixel using \texttt{grizli} software \citep{2023zndo...7712834B}.

With images from HST, JWST/NIRCam, JWST/MIRI, and ALMA, we have gathered a comprehensive data set of multi-wavelength observations for PACS-830. These observations, ranging from rest-frame UV and optical to NIR, MIR, and FIR, offer a detailed, spatially-resolved view of a high-z starburst galaxy. This approach allows us to dissect its complex morphology and physical properties. PACS-830 is similar to our study of another starburst galaxy from the same sample, PACS-819 ($11\times$ MS, as detailed in \citealt{liu_jwst_2024}), where we ruled out the major-merger scenario as the trigger for its starburst activity, despite PACS-819 having a clumpy morphology in the HST observations at rest-frame UV.

\begin{figure}
 \centering
 \includegraphics[width=6.8cm]{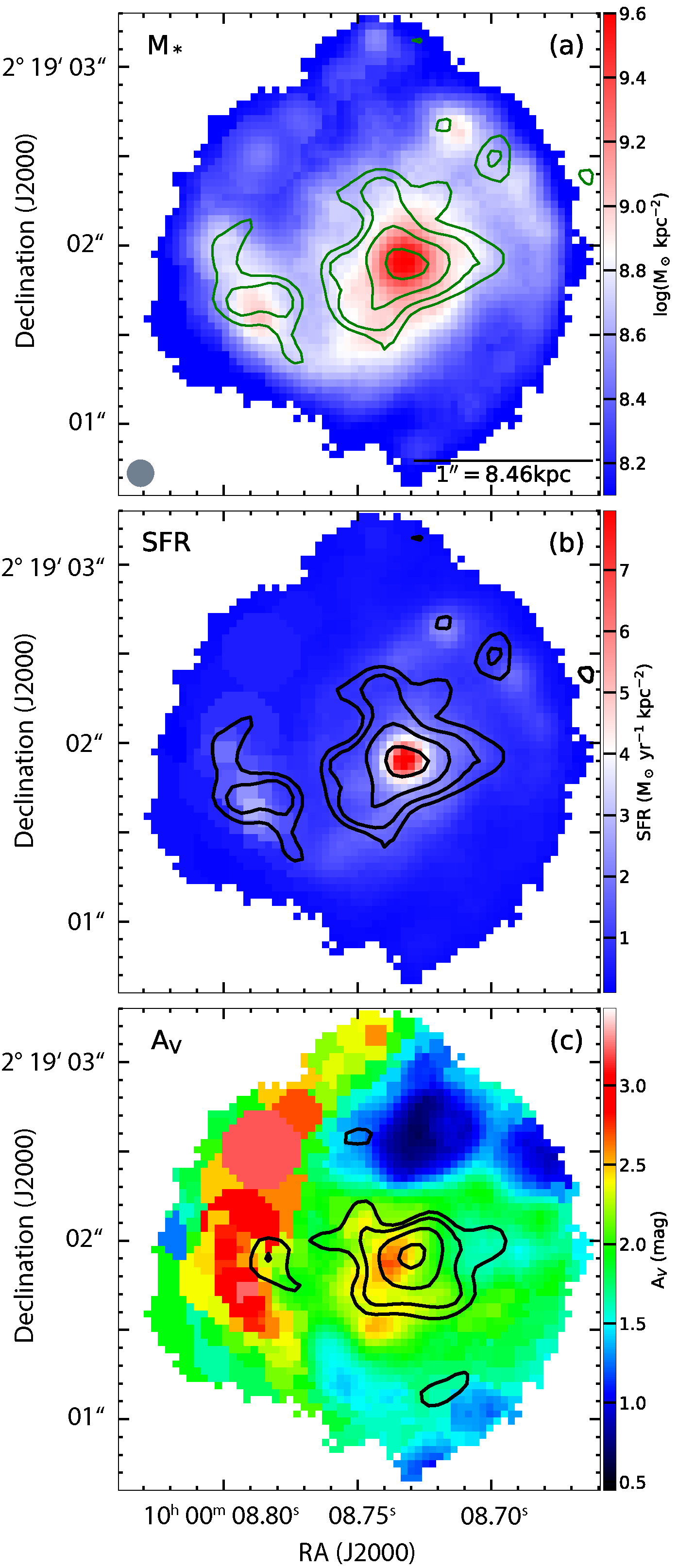}
\caption{Spatially-resolved SED fitting: stellar mass (a), SFR (b), and dust extinction (c). CO ($J$=5--4) contours are overlaid in panels a and b for comparison. In panel c, dust continuum contours are shown for comparison with the dust attenuation. Contours for the CO emission are plotted at levels of 3, 4, 6, and 9$\,\times\,\sigma_{\mathrm{rms}}$. Contours for the dust continuum are at the same levels, with $\sigma_{\mathrm{rms}} = 0.01\,\mathrm{mJy}\,\mathrm{beam}^{-1}$. The physical scale, 1$^{\prime\prime}$~=~8.46 kpc, is shown in the lower-right corner of panel a.}

\label{fig:p2p}
\end{figure}

 \begin{figure}
 \centering
 \includegraphics[width=8cm]{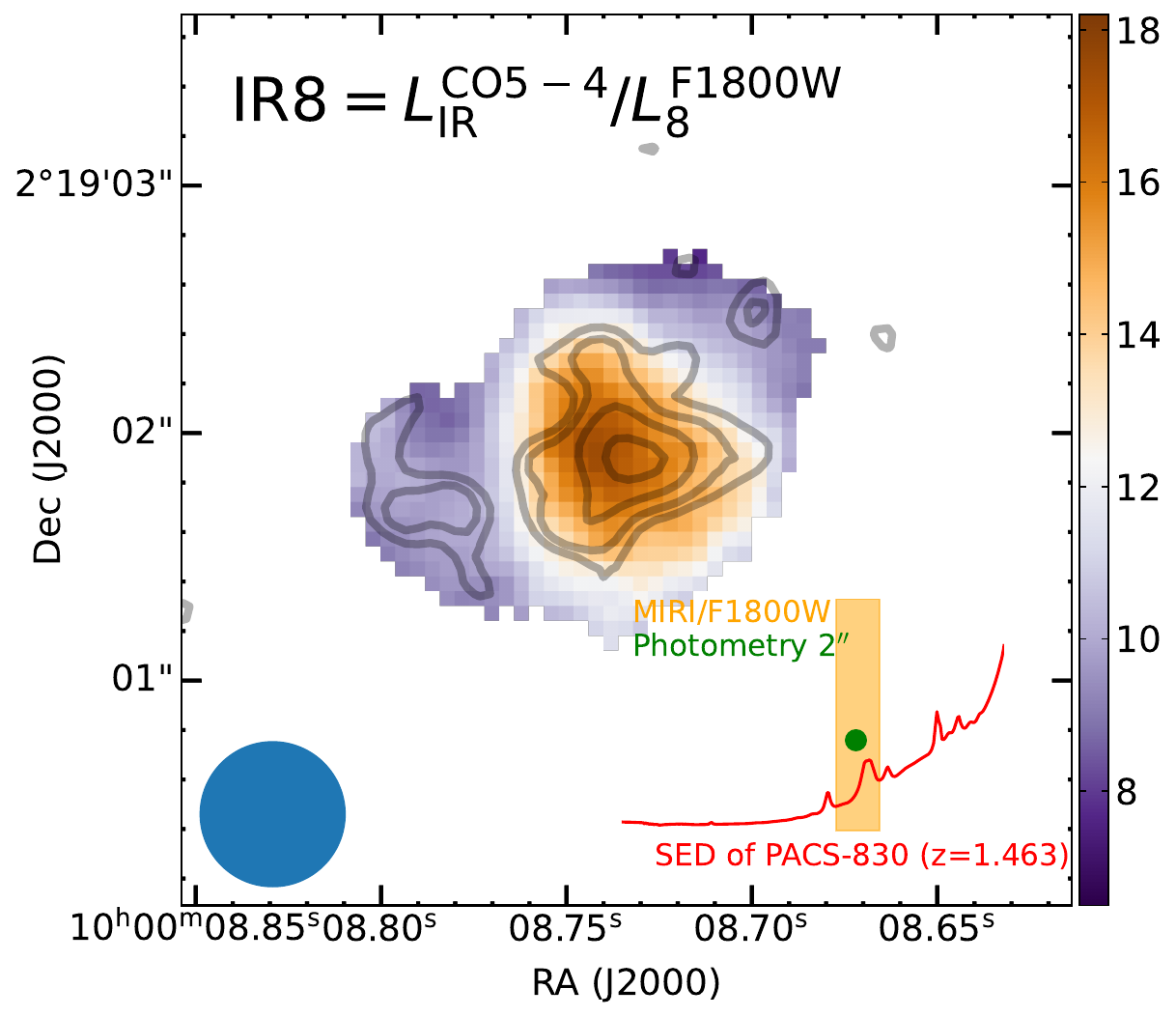}
 \caption{Spatially-resolved $IR8$ map of PACS-830. The PSF of F1800W is shown in the lower-left corner. CO J=5--4 contours are overlaid for comparison, highlighting regions of the dense molecular gas. The figure shows the same field of view as Fig.~\ref{fig:observations}. The subpanel in the lower-right corner demonstrates the modeled spectral energy distribution (SED) of the PAH emission of PACS-830, as detailed in \citet{liu_co_2021}. The F1800W aperture photometry (r=2$^{\prime\prime}$; green dot) aligns well with the SED.}
 \label{fig:$IR8$}
 \end{figure}

\section{Image Modeling}\label{sec:modeling}

\subsection{Pixel-by-pixel SED fitting} \label{sec:p2p}
Using the HST and JWST/NIRCam images, we conducted pixel-by-pixel SED fitting (Figure~\ref{fig:p2p}) to discern the stellar mass distribution and the magnitudes of the dust attenuation for this complex starburst system. 
First, we detected a foreground galaxy\footnote{Based on the Lyman break wavelength determined using GALEX, we conclude this galaxy has a redshift below 0.5, in agreement with the estimated photo-z in the COSMOS catalogs \citep{capak_first_2007,ilbert_cosmos_2009}.} to the east which slightly overlaps with PACS-830 (Figure~\ref{fig:observations}.a), next to the spiral arm.  We subtracted this foreground galaxy using \texttt{Galight} \citep{ding_mass_2020} in filters bluer than F277W, where the foreground galaxy contributes significantly. We modeled the whole field of view of F606W, where PACS-830 and the foreground galaxy have the best contrast. Then, we fixed the shape and position of the foreground galaxy when modeling the galaxies in other filters and subtracted the foreground component from the original images. The PSF of HST images are based on field stars auto-detected by \texttt{Galight} while the JWST PSFs are generated using PSFEx \citep{2011ASPC..442..435B} and provided by \citet{tanaka_m_rm_2024}. The fitting result of F606W is shown in Figure~\ref{fig:img_model_fit} as an example of this exercise.

The foreground subtracted images, F277W and F444W images are used to construct the pixel-by-pixel SED model following the same procedures as in \citet{liu_jwst_2024}. The wavelength coverage is from rest-frame UV 1650~{\AA} to near-IR 1.8~$\mu$m. We construct a binned flux cube with a spatial scale of 40 mas/pixel and FWHM matching the F444W PSF (0.16 arcsec) using \texttt{piXedfit\_bin} in \texttt{pixedfit} \citep{abdurrouf_introducing_2021}. When binning, we required all bins with similar SED shapes to have an S/N $\ge$ 3 in JWST bands. Before fitting, the bins containing the PSF spikes shown in Figure~\ref{fig:observations}.(e-f) are manually masked. When running SED fitting pixel-by-pixel with \texttt{bagpipes} \citep{carnall_inferring_2018}, we used the setting in \citet{liu_jwst_2024} designed for another starburst at a similar redshift to fit a constant star-formation history model with the solar metallicity. Figure~\ref{fig:p2p} shows the stellar mass map (a), SFR map (b), and dust attenuation map (c) from the pixel-by-pixel SED fitting. We used the F277W image as a reference to redistribute the SED-based stellar mass from each spatial bin to individual pixels, assuming that stellar mass linearly scales with the F277W flux within the same bin. This allows us to create a higher-resolution stellar mass map, leveraging the fact that the F277W band (rest-frame 1.1 $\mu$m) closely traces the stellar light distribution.

\subsection{Spatially-resolved $IR8$ map}

\citet{elbaz_goodsherschel_2011} used Spitzer photometry at the rest-frame $8\mu$m for SFGs up to $z=2.5$ to determine the contribution of PAH $7.7\mu$m emission to the total infrared luminosity. Despite the drawback of not removing the continuum, the dominant PAH $7.7\mu$m emission still established a tight empirical relationship between the luminosity at 8$\mu$m ($L_8$) and the total infrared luminosity ($L_\mathrm{IR}$).

Thanks to the high sensitivity and resolution of ALMA and JWST, we can investigate this relationship on a spatially-resolved manner. The F1800W filter covers the rest-frame wavelength of $8\mu$m of PACS-830, as shown in Figure~\ref{fig:$IR8$}. This allows us to use the flux in F1800W as a spatially resolved proxy for $L_8$ at 0.59$^{\prime\prime}$ resolution. To account for the difference between the flux in F1800W and IRAC-$8\mu \mathrm{m}$ passband, we apply a k-correction using the SED template of M82 from \citet{polletta_spectral_2007} to make a fair comparison with the results in \citet{elbaz_goodsherschel_2011}. The corrected flux is 1.3 times lower. We infer the spatially resolved $L_8$ map from the F1800W image by converting the corrected flux to luminosity. To assess the spatially resolved $L_\mathrm{IR}$, we utilize high-resolution ALMA observations of CO ($J$=5--4) instead of the dust continuum. This approach avoids the extra uncertainty introduced by temperature variations in the dust component, as CO ($J$=5--4) flux is independent of dust modeling, allowing us to infer $L_\mathrm{IR}$ in each pixel at 0.3 arcsec resolution. The conversion is based on the empirical linear relation between CO ($J$=5--4) luminosity $L'_\mathrm{CO[5-4]}$ and the total infrared luminosity $L_{\mathrm{IR}}$, well demonstrated in \citet{greve_star_2014,daddi_co_2015,liu_high-j_2015,valentino_co_2020}. We adopt the equation in \citet{daddi_co_2015}, as follows:
\begin{equation}\label{eq:1}
\log L_{\mathrm{TIR}} / L_{\odot}=\log L_{\mathrm{CO}[5-4]}^{\prime}/(\text{K}~\text{km/s}~\text{pc}^2)+2.52,
\end{equation}
where $\log L_{\mathrm{CO}[5-4]}^{\prime}$ is calculated using Eq.(1) in \citet{solomon_molecular_2005}. 

By dividing the $L_\mathrm{IR}$ map by the $L_8$ map, we derive the spatially resolved $IR8$ ($\equiv L_\mathrm{IR}/L_8$) map shown in Figure~\ref{fig:$IR8$}, which indicates the PAH contribution to the total infrared luminosity. Prior to dividing these maps, the $L_\mathrm{IR}$ map is convolved with a Gaussian kernel to match the FWHM of the PSF of F1800W (0.59 arcsec, see Sec.\ref{sec:semodel} for details), and the region with a CO S/N lower than 4 is masked. Applying the same mask on the F1800W image, the remaining pixels have a S/N higher than 30. The $L_\mathrm{IR}$ map is also regridded to the image of F1800W. The uncertainty in each pixel of the $IR8$ map is dominated by the 0.2 dex uncertainty in Eq.~\ref{eq:1}. The systematic uncertainty of imaging processing is minimal and it does not affect the conclusions in the following sections, as the astrometry offset remains below one pixel, while the F1800W point-spread function (PSF) is about 10 times larger than the pixel scale (60\,mas).

\subsection{S\'ersic model fitting}\label{sec:semodel}
Two-dimensional modeling of the galaxy morphology using S\'ersic models allows for a precise quantitative analysis of the distribution of galaxy emissions while mitigating the impact of PSF smearing (e.g., see \citealt{magnelli_ceers_2023} for attempts with MIRI images). Here, we employ S\'ersic fitting to analyze the distribution of the PAH emissions in F1800W using \texttt{Galight} \citep{2020ApJ...888...37D}.

We utilize \texttt{WebbPSF} \citep{perrin_simulating_2012} to model the PSF of F1800W, given the observation configuration (e.g., observation date, integration time, and position angle). It is an alternative method since we fail to generate an empirical PSF due to the absence of adequate stars in the field of view. We first attempted to fit the entire galaxy with a single S\'ersic component, but this resulted in significant residuals in the spiral arm regions to the east and north, which are cospatial with the red and blue regions in the  RGB Figure~\ref{fig:observations}.m. The construction of the RGB images is detailed in the following section. To better account for the galaxy’s structure, we performed model fitting with three components: a central S\'ersic to represent the unresolved core, and two additional S\'ersic components representing the spiral arms to the east and north. These arms exhibit considerable star formation activity, as traced by CO and UV emission, and they also peak in PAH emission. Modeling them separately improves the fit and reduces residuals in these regions. The model is optimized with a Particle Swarm Optimizer \citep[PSO, ][]{kennedy_particle_1995}. The best-fit S\'ersic index (n) of the central component is 0.98, with an effective radius R$_e$ of 0.49 arcsec. The effective radius R$_e$ is calculated as the half-light radius.

\subsection{Spergel model fitting}\label{sec:spmodel}
ALMA measures visibilities in the \textit{uv}-plane instead of acquiring direct images. The structure in the reconstructed images is composed of noise and Gaussians created by the cleaning process, which is affected by the weighting scheme and the configurations for cleaning. Therefore, to accurately assess the morphology of CO ($J$=5--4), we fit the visibilities of the emission line in the \textit{uv}-plane with a Spergel model \citep{spergel_analytical_2010} using \texttt{GILDAS}. In \textit{uv}-space, the Spergel model provides a good approximation of the S\'ersic profile which lacks analytical Fourier transformability \citep{tan_fitting_2024}. The fitted result for R$_e$ is 0.51$\pm$0.02 arcsec, similar to the result based on the F1800W image. The Spergel index $\nu$ is estimated to be -0.2$\pm0.1$. Using the fitted $\nu$, radius and beam size of CO ($J$=5--4), we derive a S\'ersic index of 1.7$\pm$0.2, converted from $\nu$ using the Eq.(3) in \citet{tan_fitting_2024}.

\section{Results} \label{sec:results}

\subsection{The multi-wavelength morphology of PACS-830}
The multi-wavelength observations of PACS-830 (Figure~\ref{fig:observations}) depict a morphological transition from a major-merger-like structure in the rest-frame UV to a face-on, grand-design spiral galaxy with two well-formed arms in the mid-IR. As revealed by JWST, similar wavelength-dependent changes in morphology have been seen in starbursts at cosmic noon and other dusty SFGs, due to high dust attenuation in central star-forming regions \citep[e.g.,][]{ferreira_jwst_2023,kokorev_dust_2023,bail_jwstceers_2024,gillman_structure_2024,liu_jwst_2024, faisst_cosmos-web_2024,polletta_jwsts_2024}.

To indicate variations in morphology with wavelength \citep[e.g.,][]{bail_jwstceers_2024}, we created RGB images from two sets of JWST images after subtracting the foreground galaxies. The color image in Figure~\ref{fig:observations}.l uses F814W, F115W, and F444W to indicate the blue (rest-frame UV, tracing the unobscured stars), green (rest-frame optical), and red (rest-frame near-infrared, tracing the obscured stellar population) components. The second color image (Figure~\ref{fig:observations}.m) uses F435W, F444W, and F1800W thus presenting the blue (rest-frame UV, tracing the unobscured stars), green (rest-frame near-infrared, tracing the obscured stellar population), and red (rest-frame mid-infrared, PAH emissions) components. Before combining, the images are matched in resolution to the image with the widest PSF, which are F444W (0.16 arcsec) and F1800W (0.59 arcsec) for the two sets respectively. 

As shown in Figure~\ref{fig:observations}.l, the morphology of the starburst consists of a dusty starburst core, resembling a galactic bulge, along with many bright knots distributed along two evident spiral arms. There are faint (spiral) features connecting these components which remain below the detection threshold in HST images. The RGB images (Figure~\ref{fig:observations}.l and m) reveal that the clumps, mainly along the arms, vary in color, thus likely have different ISM conditions. Among these clumps are two particularly notable regions: one to the north of the central core, which is UV bright, and another to the east, which shows strong dust attenuation. The clumps near the central core may contribute to the formation of the galaxy's bulge.

We emphasize that the CO ($J$=5--4) emission distinctly reveals the spiral arms\footnote{Complementary kinematic information and a discussion of the moment 1 and 2 maps, including a modeling attempt, are provided in Appendix~\ref{appendixB}.} , with notably clearer detection in the eastern arm. We highlight this by overlaying the CO ($J$=5--4) contours on the F444W image, which traces the stellar component (Figure~\ref{fig:observations}.g). We fit the CO ($J$=5--4) emission of the eastern arm and the core simultaneously with two Gaussian models using the Cube Analysis and Rendering Tool for Astronomy \citep[CARTA,][]{angus_comrie_2021_4905459} and find the flux contribution of the arm to the total is 21$\pm$6\%. 

The dense gas in the southeastern spiral arm appears highly extended in Figure~\ref{fig:observations}.j, which could result from two bright clumps being blended into a single structure—consistent with how the two clumps in Figure~\ref{fig:observations}.f appear merged in Figure~\ref{fig:observations}.h. To assess the impact of resolution, we degraded the F277W image to the F770W resolution via PSF matching. The resulting flux ratio map shows no clear evidence for multiple clumps in the arms or bulge, indicating that the extended appearance in F770W is largely due to resolution-induced blending.

As shown in Figure~\ref{fig:observations}.j and k, the bulk of star formation, as indicated by CO and continuum emission, is concentrated in the galactic center,  suggesting the presence of a bulge in formation, a feature commonly observed at cosmic noon \citep[e.g.,][]{jimenez-andrade_radio_2019,tan_situ_2024}. The slightly distorted morphology of the starburst core in CO($J$=5--4) may similarly arise from limited resolution and sensitivity, as clumps and bulge structures, distinguishable by color in the RGB image (Figure~\ref{fig:observations}.l), could become spatially blended.

The stellar mass map (Figure~\ref{fig:p2p}.a) clearly shows the spiral arms, where the stellar mass surface density $\Sigma_\mathrm{M_*}$ is roughly double that of the off-arm regions of the disk. In the central region, $\Sigma_{\mathrm{M}_*}$ peaks at $10^{9.6}\,\mathrm{M}_{\sun}\,\mathrm{kpc}^{-2}$—about 25 times higher than the off-arm regions and 10 times higher than the arms—indicating a bulge. Though the eastern arm is slightly more massive than the western one, indicating a slight asymmetry in the morphology, the mass distribution shows no evidence of a major merger triggering the starburst.

The SFR map (Figure~\ref{fig:p2p}) from pixel-by-pixel SED fitting based on the HST and NIRCam images shows regions of enhanced star-forming activity, i.e., the galactic core and the spiral arm to the east. These sites of enhanced SF activity are also co-spatial with the peaks in CO (J=5-4).

The $\mathrm{A}_\mathrm{V}$ map reveals that the difference in dust attenuation between the northern UV emitting part and the central part is two magnitudes, resulting in the merger-like morphology in the rest-frame UV. The part with the highest dust attenuation resides in the eastern spiral arm. It aligns with the fact that the eastern arm is much fainter in the rest-frame optical (Figure~\ref{fig:observations}.d and e) but becomes comparably bright in the rest-frame NIR (Figure~\ref{fig:observations}.g and h). However, the dust continuum does not peak here; it is in its close vicinity, as shown in Figure~\ref{fig:p2p}.b. This could simply be a result of a low signal-to-noise ratio (SNR) of the dust continuum. Consequently, we are reluctant to use the dust continuum to infer the molecular gas distribution as in \citet{liu_jwst_2024} or include it in the pixel-by-pixel SED fitting. A deeper and higher resolution observation is necessary to address this question, which is beyond the scope of the current data set. The mismatch in the core region could be an effect of convolution as, after convolving the bluer bands to the coarse PSF, the light from the blue clumps to the west of the core may contaminate the flux in the center, thus leading to a slightly lower $\mathrm{A}_\mathrm{V}$.

 \begin{figure}
 \centering
 \includegraphics[width=8cm]{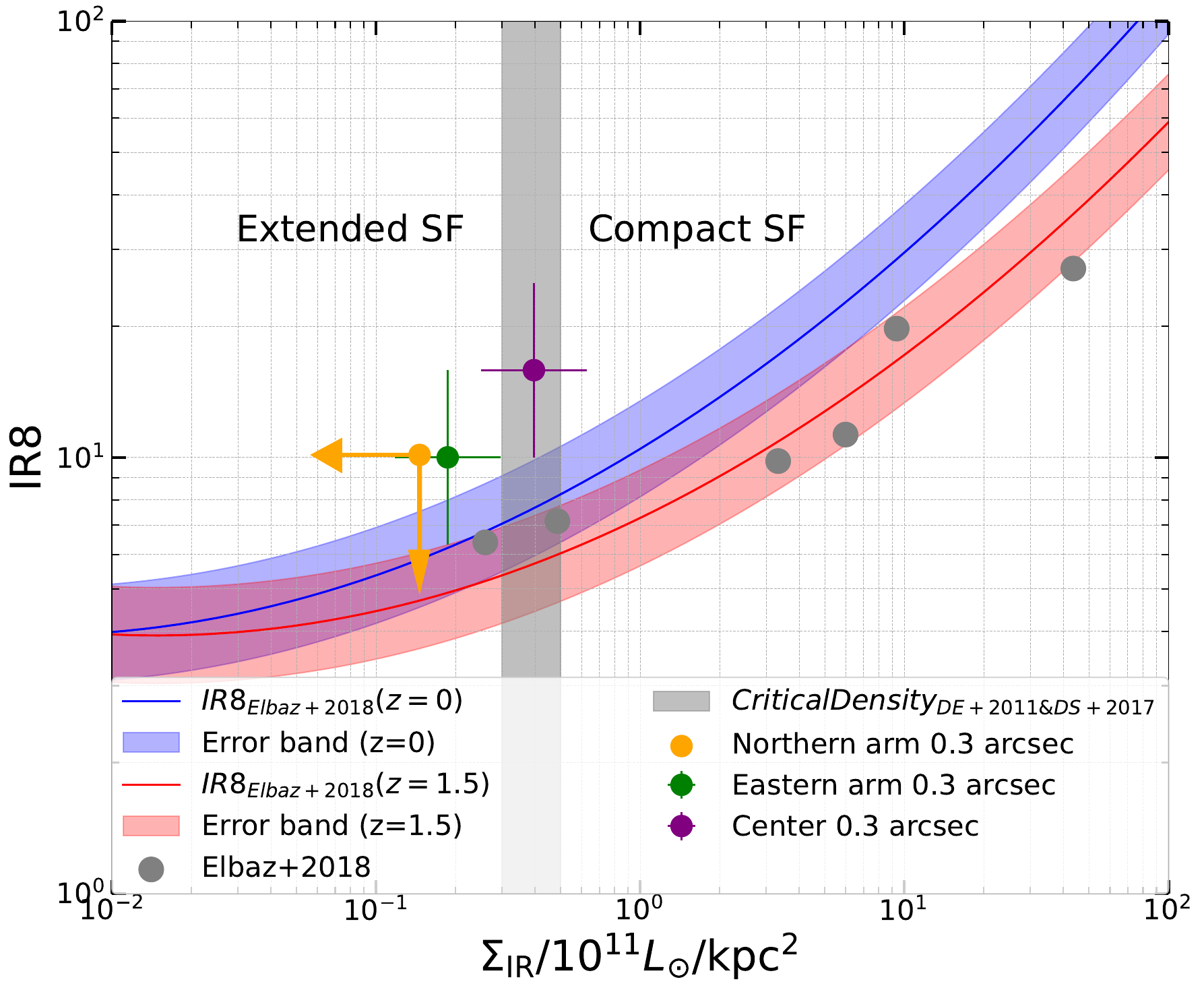}
 \caption{$IR8$ vs. $\Sigma_\mathrm{IR}$, with measurements, errorbars (0.2~dex) and upper limits from the eastern and northern spiral arms and central region ($0^{\prime\prime}.3$) shown in green, orange and purple, respectively. The 'critical density' in \citet{elbaz_goodsherschel_2011} and \citet{diaz-santos_herschelpacs_2017} is highlighted by the gray shaded region. Two fitted lines at $z=0$ and $z=1.5$ from \citet{elbaz_starbursts_2018} are shown in blue and red, with shaded regions representing a 0.11 dex error range. The grey data points are the starburst galaxies at $1.5<z<2.5$ from \citet{elbaz_starbursts_2018}.}
 \label{fig:$sigmair$}
 \end{figure}

\subsection{An $IR8$ enhancement in the core: A PAH deficit in the starbursting core?}
$IR8$ is defined as the ratio of the infrared luminosity ($L_\mathrm{IR}$), to the luminosity at 8 $\mu$m, $L_8$, which \citet{elbaz_goodsherschel_2011} found to closely correlate with the surface density of MIR and FIR luminosity. In our case, we utilize the CO ($J$=5--4) line as a proxy for $L_\mathrm{IR}$ and the flux in F1800W as $L_8$. The uncertainty of $IR8$ is mainly contributed by the conversion from the luminosity of CO (J=5--4) to between $L_\mathrm{IR}$, which is a small scatter of $\sim$0.2 dex. In Figure~\ref{fig:$IR8$}, the $IR8$ map peaks at the center, with a value around 18, which is roughly double the values observed in the surrounding spiral arms, as indicated by the CO contours. While we acknowledge that our measured $IR8$ may be affected by systematic uncertainties related to CO excitation or temperature variations across different regions of the galaxy, the F1800W data provide a distinct morphological signature: the southeastern spiral arm exhibits a peak flux approximately 80\% that of the central region. In contrast, the CO($J$=5--4) emission and the dust continuum show more centrally concentrated morphologies with weaker emission in the southeastern arm. This discrepancy suggests that systematic uncertainties alone are unlikely to account for the pronounced $IR8$ variations.

We reiterate that there is no evidence suggesting a significant AGN influence. While we cannot completely rule out the presence of a dust-obscured, relatively faint AGN, its impact on both the spectroscopy and SED appears negligible (Sec.~\ref{sec:data}) and cannot account for the strong variation in the IR8 map. As shown in previous studies using Spitzer spectroscopy \citep[e.g.,][]{desai_pah_2007}, strong dust continuum emission, particularly from compact nuclear regions possibly associated with AGN, can dilute PAH features and mimic a deficit in spectroscopy. However, if such a hidden AGN were present at the core, we would expect enhanced F1800W emission due to a mid-IR continuum excess \citep[e.g., ][]{donley_spitzers_2008,casey_far-infrared_2012}, even if PAH features were suppressed. Instead, we observe the opposite: the F1800W emission is fainter.

The $IR8$ enhancement is further supported by our 2D modeling using S\'ersic and Spergel profiles. While both models estimate the radius of the central region to be about 0.5 arcseconds (4.3~kpc), the S\'ersic index derived from the F1800W map (n$\sim$1) is lower than that from the CO map (n$\sim$1.7). This suggests that the PAH distribution is more disk-like, while $L_\mathrm{IR}$ traced by the CO map has a more concentrated or compact distribution. Therefore, one simple explanation for this elevated $IR8$ value and disk-like PAH distribution is a deficit in PAH, possibly caused by the destruction of PAH molecules due to the hard radiation field by intense star formation activity in the starburst core \citep[e.g., ][]{draine_infrared_2007}.

However, some observations find no obvious correlation between the hardness of the radiation field or the emission-line width and IR8 \citep{inami_mid-infrared_2013,stierwalt_mid-infrared_2014}. Alternatively, the $IR8$ difference across various star-forming regions may be linked to dust distribution and temperature, as explained by the concept of "dust-bounded" regions \citep{abel_dust-bounded_2009,diaz-santos_herschelpacs_2017}. Dust can absorb a fraction of the ionizing radiation, preventing it from reaching the photodissociation region (PDR). This absorption results in an increase in dust temperature and a decrease in PAH emission from PDRs. In the relation between $IR8$ and $\Sigma_\mathrm{IR}$, $IR8$ remains relatively flat with respect to the radiation field indicator $\Sigma_\mathrm{IR}$ until it reaches a critical density, $\Sigma_\mathrm{IR}^\mathrm{crit}\sim 3-5\times10^{10}~L_\odot~\mathrm{kpc}^{-2}$, as shown in Figure~\ref{fig:$sigmair$}. Beyond this threshold, $IR8$ increases with $\Sigma_\mathrm{IR}$, a trend confirmed by observations from \citet{elbaz_starbursts_2018}. In PACS-830, we apply three r=0.$^{\prime\prime}3$ apertures to measure the $\Sigma_\mathrm{IR}$ in the eastern and northern spiral arm and starbursting core from the original CO ($J$=5--4) map, respectively. We note that, since the northern arm is not robustly detected in CO ($J$=5--4), we treat the CO luminosity as an upper limit. Consequently, the derived values of IR8 and $\Sigma_\mathrm{IR}$ for this region should also be considered upper limits. As depicted in Figure~\ref{fig:$sigmair$}, the result shows that the spiral arm regions are at a lower density ($\Sigma_\mathrm{IR}\sim2\times10^{10}~L_\odot~\mathrm{kpc}^{-2}$), while the core region is of a higher density, at the critical densities ($\Sigma_\mathrm{IR}\sim4\times10^{10}~L_\odot~\mathrm{kpc}^{-2}$), corresponding to a higher $IR8$. The core density is not exceptionally high---i.e., it does not significantly exceed the critical density—as expected for a moderate starburst ($4\times$ MS) with a radius of 0.5$^{\prime\prime}$, as derived from 2D modeling with JWST and ALMA \citep[cf. compact submillimeter sources;][]{puglisi_main_2019, puglisi_sub-millimetre_2021, tan_situ_2024}. However, the density remains sufficient to reveal distinct spatial variations in $IR8$, consistent with theoretical expectations and prior observations on global properties \citep{elbaz_starbursts_2018}.

 \begin{figure}
 \centering
 \includegraphics[width=8cm]{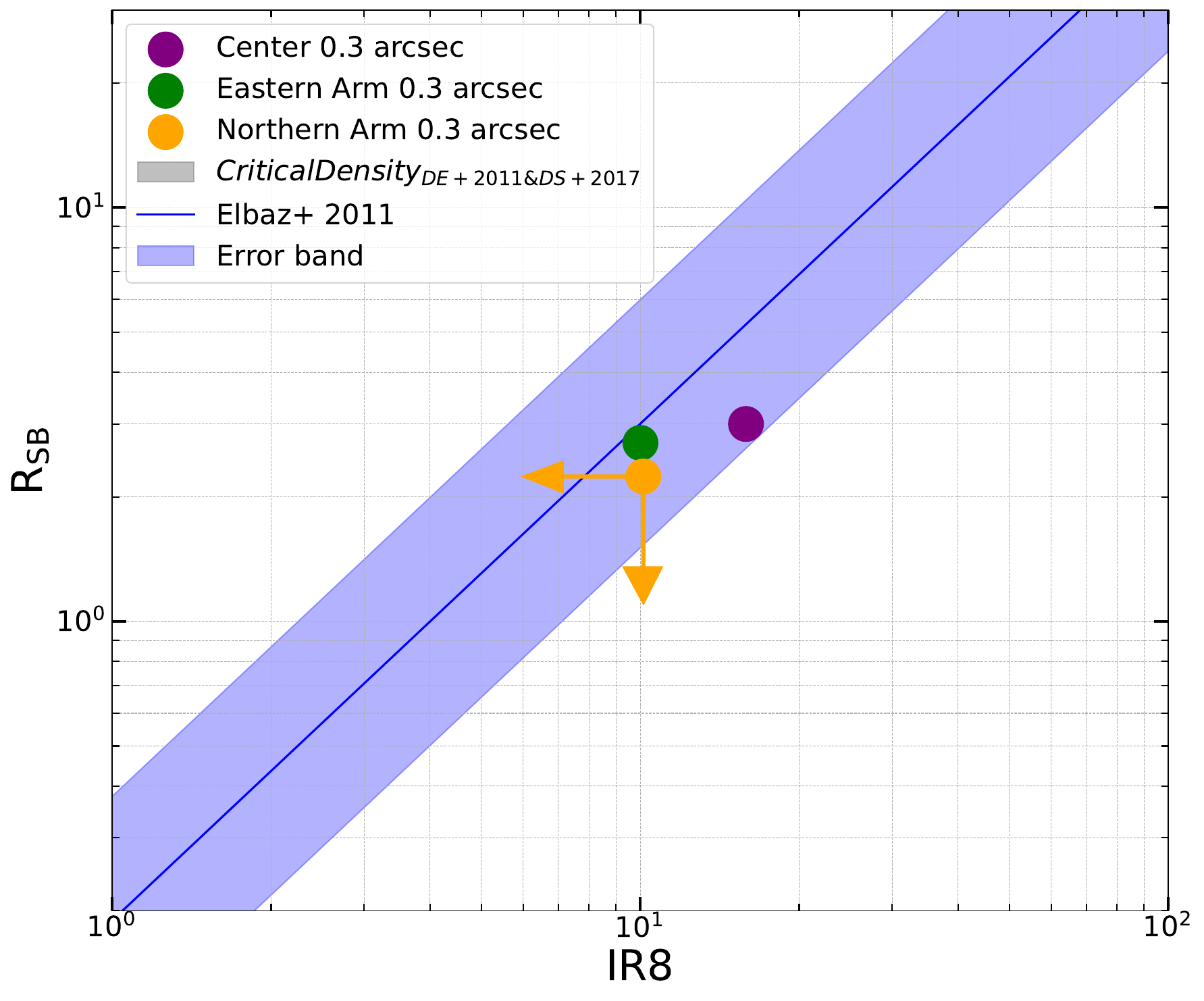}
 \caption{$R_\mathrm{SB}$ vs. $IR8$, with measurements and upper limits from the eastern and northern spiral arms and central region ($0^{\prime\prime}.3$) shown in green, orange, and purple, respectively. One fitted line of local galaxies from \citet{elbaz_goodsherschel_2011} is shown in blue, with shaded regions representing a 0.3 dex dispersion.}
 \label{fig:rsb}
 \end{figure}

For local galaxies, $IR8$ also indicates the starburstiness $R_\mathrm{SB}$, i.e., the distance above the MS. Therefore, we examine $R_\mathrm{SB}$ in the core region and spiral arms using the SFR (i.e., CO ($J$=5--4)) and stellar mass maps. The SFR is further inferred from the CO ($J$=5--4) converted $L_\mathrm{TIR}$ map based on the calibration by \citet{hao_dust-corrected_2011} and \citet{murphy_calibrating_2011}. Since the CO ($J$=5--4) map has a coarser resolution, we first smooth the stellar mass map by doubling the FWHM of its PSF (0.32$^{\prime\prime}$) and then convolve the CO ($J$=5--4) map to match this new Gaussian-like PSF. We then apply the same apertures as in Figure~\ref{fig:$sigmair$} to measure the $R_\mathrm{SB}$ based on the MS of \citet{speagle_highly_2014}. The results well align with the relation in \citet{elbaz_goodsherschel_2011}, shown in Figure~\ref{fig:rsb}, in contrast to the offset in Figure~\ref{fig:$sigmair$}. This indicates that, for a given $R_\mathrm{SB}$, $\Sigma_\mathrm{IR}$ is lower than expected, suggesting that the star formation mode in this spiral-starburst galaxy may differ from the compact starbursts triggered by major mergers.

\section{Discussion: Formation and Impact of Substructures in PACS-830}\label{sec:discussion}

In the pre-JWST era, \citet{2014ApJ...781...11E} suggested a morphological evolutionary path for star-forming galaxies, commonly progressing from highly turbulent, gas-rich disks with giant clumps to more ordered spiral galaxies based on HST observations in the rest-frame optical. However, our JWST and ALMA observations of PACS-830 reveal a rare mixed case, in which clumps, spiral arms, and a bulge in formation coexist in a single system, aligning with other emerging JWST results \citep{kuhn_jwst_2024,kalita_multi-wavelength_2025}. Notably, we observe a gradual change from a clumpy morphology in the rest-frame UV to a more spiral-like appearance at NIR and MIR wavelengths, hinting that previous conclusions may have been influenced by both high dust attenuation in distant star-forming galaxies and the limitations of previous-generation instruments like HST.

In this section, we connect our key findings to existing literature and explore how these substructures emerge and evolve, focusing on the implications of clumps, spiral arms, and bulges for the physical properties across the galaxy. We also highlight the associated star formation rates and PAH characteristics as potential diagnostics of the varying conditions within different substructures, thereby providing insights into the interplay of morphological evolution and star formation in distant starburst systems. In particular, we detect a pronounced PAH deficit in the forming bulge, potentially attributable to PAH destruction by the intense radiation field or decreased PDR emissions at the galactic center---an effect likely linked to the enhanced star-forming activity therein sustained by the evolving structure of PACS-830.

\subsection{A starbursting bulge}
Triggered either by violent disk instabilities (VDI; \citealt{dekel_cold_2009}) or minor-interactions with a neighboring galaxy (see Sec.~\ref{subsec:spiral}), the fragmented clumps as well as the spiral arms can contribute to disk instabilities within PACS-830 thus driving bulge formation \citep[see][for a review]{bournaud_bulge_2016}. Over time, the clumps along the spiral arms possibly migrate inward due to gravitational torques and dynamical friction. Additionally, the presence of spiral arms and clumps points to gas inflow, as these non-axisymmetric structures transfer the angular momentum and allow gas to migrate inward. Therefore, the bulge formation may be fueled by a combination of gas and stars, consistent with our observations: the irregular CO shape and the UV/optical clumps spotted in the core region may indicate that the central bulge is growing by both in-situ and ex-situ star formation.

The evidence of a deficit in PAH emission may indicate rapid bulge formation due to a higher star formation efficiency (SFE). \citet{cortzen_pahs_2019} report that PAH 6.2$~\mu$m correlates with CO (J=1--0) from $z=0-4$ for SMGs, and \citet{shivaei_tight_2024} investigate the relationship between PAH 7.7$~\mu$m and CO (J=1--0) in 14 MS galaxies at $z=1-3$, reinforcing that PAH emissions can be used as a proxy for molecular gas mass. A higher $IR8$ value in PACS-830, therefore, indicates a higher SFE, since SFE is proportional to $IR8$ (SFE = SFR/M$_\mathrm{gas}\propto$$IR8$). Such a high SFE in the core region aligns with the rapid bulge formation indicated by the morphology with high global instabilities that may quench the starburst after depleting the central gas rapidly, supporting the inside-out growth scenario. Similar cases are also found in z$\sim$2 lensed, massive starburst galaxies at hundreds of parsec scales, but with more centrally depleted molecular gas \citep{liu_600_2023,liu_detailed_2024}, possibly linking to different stages of inside-out quenching.

However, it’s important to highlight a limitation in our analysis: $IR8$ is just a color index, and not a direct comparison of PAH 7.7$~\mu$m and CO ($J$=5--4) emissions. Although we infer that the PAH 7.7$~\mu$m emission is strong and broad, based on the integrated SED fitting from \citet{liu_co_2021} and our new MIRI/F1800W photometry (subplot in Fig.~\ref{fig:$IR8$}), we remain uncertain about whether the contribution from the underlying continuum remains consistent. Nevertheless, continuum dilution is unlikely to explain the observed flux deficit at 8~$\mu$m. To solidify our conclusions, a further spectroscopic study of PACS-830 and a larger sample are necessary.

Regardless of the interpretation, we conclude that this $IR8$ enhancement could be a signature of an intense star-forming region, possibly linked to a rapidly growing bulge. Spatially comparing PAH tracers with indicators of star formation (e.g., CO high-J transitions or Paschen lines), derived from different physical processes, could be a useful exercise to hunt for SBs hidden in MS \citep{elbaz_starbursts_2018} or bulges in formation. This approach is especially promising given the abundant ALMA archival data, the gradually building-up JWST/NIRCam WFSS archival data \citep[e.g., COSMOS-3D,][]{kakiichi_cosmos-3d_2024} on infrared emission lines and MIRI broadband imagery.

\subsection{Clumps along the spiral arms}
Giant clumps can disrupt the symmetry of the overall mass distribution, thereby enhancing star formation across the disk, potentially seen in the bright CO clump detected in the eastern spiral arm. These clumps are thought to form through disk fragmentation \citep{elmegreen_bulge_2008}, a scenario supported by kinematic observations of clumps in massive gas-rich disks \citep{genzel_sins_2011}. In PACS-830, the gas-to-stellar mass ratio ($\sim$1) of the entire galaxy meets the criterion for a gas-rich disk \citep{silverman_concurrent_2018}; thus, fragmentation could be triggered by a secular process, i.e., VDI, as expected in \citet{faisst_cosmos-web_2024}.

However, the most prominent clumps lie along the spiral arms rather than being randomly scattered, hinting that gravitational interactions or other dynamical processes may also play a role (see next subsection). Despite their seemingly orderly positioning, the clumps vary in size, mass, and star formation rates. Interestingly, while the CO ($J$=5--4) and PAH emissions differ between the eastern and northwestern clumps, their IR8 values remain similar. Of different scales mentioned above, the clumps do not exhibit large differences in their radiation fields or overall starburst activity. This finding aligns with results from local (U)LIRGs in the GOALS sample, where the mid-infrared spectra of extended star-forming regions beyond $\sim$1.5 kpc from the nucleus are remarkably homogeneous, and variations in integrated MIR properties primarily reflect the diversity of nuclear conditions \citep[e.g.,][]{diaz-santos_spatial_2011,stierwalt_mid-infrared_2014}.

\subsection{Grand-design spiral arms}\label{subsec:spiral}
PACS-830 has exactly two prominent and symmetric arms along with clumps mostly located on the arms, which resemble a 'grand-design' morphology. Simulations suggest that grand design arms form due to the gravitational interactions with other galaxies instead of VDI, which usually results in multi-armed or flocculent (patchy) morphology \citep{dobbs_simulations_2010}. PACS-830 could be a past minor merger as in PACS-819 \citep{liu_jwst_2024}. Or, it is an ongoing mini merger with a neighboring candidate with a similar robust photometric redshift ($z=1.4764$, $\sigma_z=0.009$) within proximity ($<3^{\prime\prime}$) \footnote{Another neighboring galaxy ($\sim8^{\prime\prime}$ away) with a similar photo-z does not meet the selection standard in the ZFOURGE catalog \citep{straatman_fourstar_2016}.}. Its redshift is derived from the ZFOURGE survey \citep{straatman_fourstar_2016}, estimated with the photometry in six filters of the FourStar near-infrared camera on the Magellan telescope, complemented by 25 ancillary photometric filters in the COSMOS field. The color of the galaxy similar to PACS-830's also indicates that they might be a pair of interacting galaxies. 

A stellar mass of $10^{9.5}M_\odot$ and a SFR of $\sim$7 locates it as a MS galaxy. The mass ratio of it and PACS-830 is about 1/30, which is below the threshold for a minor merger \citep[e.g., 1/10 in][]{bottrell_illustristng_2024}. \citet{bottrell_illustristng_2024} highlights that mini mergers can trigger small enhancements in SFRs and asymmetries, aligning with the moderate starburst and non-axisymmetric morphology of PACS-830. Given that PACS-830 shows only a moderate offset from the MS, being a factor of 4 above the relation, the interaction of this mini merger could trigger the SFR enhancement and the formation of the grand-design spiral arms. However, this galaxy is undetected in all ALMA maps and shows only a marginal detection with S/N $\sim$4 in F1800W, preventing any further analysis of its PAH and CO emissions.

\section{Final remarks}\label{sec:summary}
With the help of JWST/MIRI, we can now spatially resolve emissions from critical dust grains—PAHs—in SFGs at cosmic noon. This opens a new window to map star formation in the mid-infrared, particularly for the diverse substructures in these galaxies revealed by JWST/NIRCam. We showcase our analysis of the starburst system PACS-830 at $z=1.463$ focusing on its compact nuclear starburst and extended regions using multi-wavelength data from HST, JWST, and ALMA. Unlike major-merger-like starbursts, PACS-830 resembles a grand-design spiral galaxy with a bulge-forming nucleus and multiple clumps along its spiral arms, as revealed by NIRCam images. Through pixel-by-pixel SED fitting, we confirm that PACS-830 is indeed a spiral galaxy rather than a multiple major merger suggested by HST images, based on its stellar distribution. This finding downplays the role of major-merger-triggered starbursts in gas-rich galaxies at cosmic noon and favors interaction-induced or VDI-induced scenarios or past minor merger events as in PACS-819 \citep{liu_jwst_2024}.

The CO ($J$=5--4) and PAH (detected by MIRI/F1800W) emission maps align well with the morphology observed in JWST/NIRCam, tracing star-formation in both the central core and the spiral arms. Notably, one of the spiral arms contributes $\sim$20\% of the total flux of CO ($J$=5--4).
The comparison of these two maps, along with 2D modeling, reveals a significant PAH deficit in the center of the galaxy. This deficit is likely due to the destruction of PAHs or decreased PDR emission caused by the intense star formation in the galactic center. Our findings reveal, for the first time, spatial variations in the $L_8$ to $L_\mathrm{IR}$ ratio, aligning with theoretical predictions and previous global measurements. Consequently, we express caution in using PAHs, particularly MIRI-based PAH indicators, to determine SFRs in a spatially resolved context.

Our findings underscore the importance of high-resolution, multi-wavelength observations in revealing the true structure and internal processes of distant galaxies. We also demonstrate that the PAH deficit could serve as a tracer for distant extreme/compact star-forming regions with the help of NIRCam WFSS and MIRI. Further, more detailed studies with enhanced resolution and broader dynamic range will be crucial for understanding the mechanisms driving star formation and galaxy evolution in the early universe.

\section*{Acknowledgements}

We thank the anonymous referee for the insightful comments to improve the manuscript. Z.L. sincerely thanks Luis C. Ho, Ryota Ikeda, Hanae Inami, Benjamin Magnelli, and Naoki Yoshida for the scientific discussions. 

Kavli IPMU was established by World Premier International Research Center Initiative (WPI), MEXT, Japan. Z.L. is supported by the Global Science Graduate Course (GSGC) program of the University of Tokyo. This research was supported by a grant from the Hayakawa Satio Fund
awarded by the Astronomical Society of Japan. J.S. is supported by JSPS KAKENHI (JP22H01262) and the World Premier International Research Center Initiative (WPI), MEXT, Japan. This work was supported by JSPS Core-to-Core Program (grant number: JPJSCCA20210003). A.P. is supported by an Anniversary Fellowship at University of Southampton.

Some of the data products presented herein were retrieved from the Dawn JWST Archive (DJA). DJA is an initiative of the Cosmic Dawn Center, which is funded by the Danish National Research Foundation under grant No. 140.
%%%%%%%%%%%%%%%%%%%%%%%%%%%%%%%%%%%%%%%%%%%%%%%%%%
\section*{Data Availability}
Some of the data presented in this article were obtained from the Mikulski Archive for Space Telescopes (MAST) at the Space Telescope Science Institute. The specific observations analyzed can be accessed via DOI: 10.17909/dny0-c072. This paper makes use of the following ALMA data: ADS/JAO.ALMA\#2016.1.01426.S. ALMA is a partnership of ESO (representing its member states), NSF (USA) and NINS (Japan), together with NRC (Canada), MOST and ASIAA (Taiwan), and KASI (Republic of Korea), in cooperation with the Republic of Chile. The Joint ALMA Observatory is operated by ESO, AUI/NRAO and NAOJ.



\bibliographystyle{mnras}
\bibliography{830} % if your bibtex file is called example.bib

\begin{thebibliography}{}
\makeatletter
\relax
\def\mn@urlcharsother{\let\do\@makeother \do\$\do\&\do\#\do\^\do\_\do\%\do\~}
\def\mn@doi{\begingroup\mn@urlcharsother \@ifnextchar [ {\mn@doi@} {\mn@doi@[]}}
\def\mn@doi@[#1]#2{\def\@tempa{#1}\ifx\@tempa\@empty \href {http://dx.doi.org/#2} {doi:#2}\else \href {http://dx.doi.org/#2} {#1}\fi \endgroup}
\def\mn@eprint#1#2{\mn@eprint@#1:#2::\@nil}
\def\mn@eprint@arXiv#1{\href {http://arxiv.org/abs/#1} {{\tt arXiv:#1}}}
\def\mn@eprint@dblp#1{\href {http://dblp.uni-trier.de/rec/bibtex/#1.xml} {dblp:#1}}
\def\mn@eprint@#1:#2:#3:#4\@nil{\def\@tempa {#1}\def\@tempb {#2}\def\@tempc {#3}\ifx \@tempc \@empty \let \@tempc \@tempb \let \@tempb \@tempa \fi \ifx \@tempb \@empty \def\@tempb {arXiv}\fi \@ifundefined {mn@eprint@\@tempb}{\@tempb:\@tempc}{\expandafter \expandafter \csname mn@eprint@\@tempb\endcsname \expandafter{\@tempc}}}

\bibitem[\protect\citeauthoryear{{Abdurro’uf}, Lin, Wu  \& Akiyama}{{Abdurro’uf} et~al.}{2021}]{abdurrouf_introducing_2021}
{Abdurro’uf} Lin Y.-T.,  Wu P.-F.,   Akiyama M.,  2021, \mn@doi [ApJS] {10.3847/1538-4365/abebe2}, 254, 15

\bibitem[\protect\citeauthoryear{Abel, Dudley, Fischer, Satyapal  \& van Hoof}{Abel et~al.}{2009}]{abel_dust-bounded_2009}
Abel N.~P.,  Dudley C.,  Fischer J.,  Satyapal S.,   van Hoof P. A.~M.,  2009, \mn@doi [ApJ] {10.1088/0004-637X/701/2/1147}, 701, 1147

\bibitem[\protect\citeauthoryear{Akins et~al.,}{Akins et~al.}{2022}]{akins_alma_2022}
Akins H.~B.,  et~al., 2022, \mn@doi [ApJ] {10.3847/1538-4357/ac795b}, 934, 64

\bibitem[\protect\citeauthoryear{Alberts et~al.,}{Alberts et~al.}{2024}]{alberts_smiles_2024}
Alberts S.,  et~al., 2024, \mn@doi [ApJ] {10.3847/1538-4357/ad7396}, 976, 224

\bibitem[\protect\citeauthoryear{Bail et~al.,}{Bail et~al.}{2024}]{bail_jwstceers_2024}
Bail A.~L.,  et~al., 2024, \mn@doi [A\&A] {10.1051/0004-6361/202347465}, 688, A53

\bibitem[\protect\citeauthoryear{Battisti, Shivaei, Park, Decleir, Calzetti, Mathew, Wisnioski  \& da Cunha}{Battisti et~al.}{2025}]{battisti_constraining_2025}
Battisti A.,  Shivaei I.,  Park H.~J.,  Decleir M.,  Calzetti D.,  Mathew J.,  Wisnioski E.,   da Cunha E.,  2025, \mn@doi [Publications of the Astronomical Society of Australia] {10.1017/pasa.2024.129}, 42, e022

\bibitem[\protect\citeauthoryear{{Bertin}}{{Bertin}}{2011}]{2011ASPC..442..435B}
{Bertin} E.,  2011, in {Evans} I.~N.,  {Accomazzi} A.,  {Mink} D.~J.,   {Rots} A.~H.,  eds,  Astronomical Society of the Pacific Conference Series Vol. 442, Astronomical Data Analysis Software and Systems XX. p.~435

\bibitem[\protect\citeauthoryear{Boquien, Burgarella, Roehlly, Buat, Ciesla, Corre, Inoue  \& Salas}{Boquien et~al.}{2019}]{boquien_cigale_2019}
Boquien M.,  Burgarella D.,  Roehlly Y.,  Buat V.,  Ciesla L.,  Corre D.,  Inoue A.~K.,   Salas H.,  2019, \mn@doi [A\&A] {10.1051/0004-6361/201834156}, 622, A103

\bibitem[\protect\citeauthoryear{Bottrell et~al.,}{Bottrell et~al.}{2024}]{bottrell_illustristng_2024}
Bottrell C.,  et~al., 2024, \mn@doi [MNRAS] {10.1093/mnras/stad2971}, 527, 6506

\bibitem[\protect\citeauthoryear{Bournaud}{Bournaud}{2016}]{bournaud_bulge_2016}
Bournaud F.,  2016, in Laurikainen E.,  Peletier R.,   Gadotti D.,  eds, , Galactic {Bulges}.
Springer International Publishing, Cham, pp 355--390, \mn@doi{10.1007/978-3-319-19378-6_13}, \url {https://doi.org/10.1007/978-3-319-19378-6_13}

\bibitem[\protect\citeauthoryear{{Brammer}}{{Brammer}}{2023}]{2023zndo...7712834B}
{Brammer} G.,  2023, {grizli}, Zenodo, \mn@doi{10.5281/zenodo.7712834}

\bibitem[\protect\citeauthoryear{{Briggs}}{{Briggs}}{1995}]{1995AAS...18711202B}
{Briggs} D.~S.,  1995, in American Astronomical Society Meeting Abstracts. p. 112.02

\bibitem[\protect\citeauthoryear{Calabrò et~al.,}{Calabrò et~al.}{2024}]{calabro_evolution_2024}
Calabrò A.,  et~al., 2024, \mn@doi [A\&A] {10.1051/0004-6361/202449768}, 690, A290

\bibitem[\protect\citeauthoryear{Calzetti}{Calzetti}{2013}]{calzetti_star_2013}
Calzetti D.,  2013, Star {Formation} {Rate} {Indicators}, \mn@doi{10.48550/arXiv.1208.2997.
}, \url {https://ui.adsabs.harvard.edu/abs/2013seg..book..419C}

\bibitem[\protect\citeauthoryear{Capak et~al.,}{Capak et~al.}{2007}]{capak_first_2007}
Capak P.,  et~al., 2007, \mn@doi [ApJS] {10.1086/519081}, 172, 99

\bibitem[\protect\citeauthoryear{Cappelluti et~al.,}{Cappelluti et~al.}{2009}]{cappelluti_xmm-newton_2009}
Cappelluti N.,  et~al., 2009, \mn@doi [A\&A] {10.1051/0004-6361/200810794}, 497, 635

\bibitem[\protect\citeauthoryear{Carnall, McLure, Dunlop  \& Dav\'e}{Carnall et~al.}{2018}]{carnall_inferring_2018}
Carnall A.~C.,  McLure R.~J.,  Dunlop J.~S.,   Dav\'e R.,  2018, \mn@doi [MNRAS] {10.1093/mnras/sty2169}, 480, 4379

\bibitem[\protect\citeauthoryear{Casey}{Casey}{2012}]{casey_far-infrared_2012}
Casey C.~M.,  2012, \mn@doi [MNRAS] {10.1111/j.1365-2966.2012.21455.x}, 425, 3094

\bibitem[\protect\citeauthoryear{Casey et~al.,}{Casey et~al.}{2023}]{casey_cosmos-web_2023}
Casey C.~M.,  et~al., 2023, \mn@doi [ApJ] {10.3847/1538-4357/acc2bc}, 954, 31

\bibitem[\protect\citeauthoryear{Chabrier}{Chabrier}{2003}]{chabrier_galactic_2003}
Chabrier G.,  2003, \mn@doi [PASP] {10.1086/376392}, 115, 763

\bibitem[\protect\citeauthoryear{Chastenet et~al.,}{Chastenet et~al.}{2023}]{chastenet_phangs-jwst_2023}
Chastenet J.,  et~al., 2023, \mn@doi [ApJ] {10.3847/2041-8213/acac94}, 944, L12

\bibitem[\protect\citeauthoryear{Civano et~al.,}{Civano et~al.}{2016}]{civano_chandra_2016}
Civano F.,  et~al., 2016, \mn@doi [ApJ] {10.3847/0004-637X/819/1/62}, 819, 62

\bibitem[\protect\citeauthoryear{Comrie et~al.,}{Comrie et~al.}{2021}]{angus_comrie_2021_4905459}
Comrie A.,  et~al., 2021, {CARTA: The Cube Analysis and Rendering Tool for Astronomy}, \mn@doi{10.5281/zenodo.4905459}, \url {https://doi.org/10.5281/zenodo.4905459}

\bibitem[\protect\citeauthoryear{Condon, Huang, Yin  \& Thuan}{Condon et~al.}{1991}]{condon_compact_1991}
Condon J.~J.,  Huang Z.~P.,  Yin Q.~F.,   Thuan T.~X.,  1991, \mn@doi [ApJ] {10.1086/170407}, 378, 65

\bibitem[\protect\citeauthoryear{{Conway}, {Cornwell}  \& {Wilkinson}}{{Conway} et~al.}{1990}]{1990MNRAS.246..490C}
{Conway} J.~E.,  {Cornwell} T.~J.,   {Wilkinson} P.~N.,  1990, \mnras, \href {https://ui.adsabs.harvard.edu/abs/1990MNRAS.246..490C} {246, 490}

\bibitem[\protect\citeauthoryear{Cornwell}{Cornwell}{2008}]{cornwell_multiscale_2008}
Cornwell T.~J.,  2008, \mn@doi [IEEE Journal of Selected Topics in Signal Processing] {10.1109/JSTSP.2008.2006388}, 2, 793

\bibitem[\protect\citeauthoryear{Cortzen et~al.,}{Cortzen et~al.}{2019}]{cortzen_pahs_2019}
Cortzen I.,  et~al., 2019, \mn@doi [MNRAS] {10.1093/mnras/sty2777}, 482, 1618

\bibitem[\protect\citeauthoryear{Cowie, González-López, Barger, Bauer, Hsu  \& Wang}{Cowie et~al.}{2018}]{cowie_submillimeter_2018}
Cowie L.~L.,  González-López J.,  Barger A.~J.,  Bauer F.~E.,  Hsu L.~Y.,   Wang W.~H.,  2018, \mn@doi [ApJ] {10.3847/1538-4357/aadc63}, 865, 106

\bibitem[\protect\citeauthoryear{Daddi et~al.,}{Daddi et~al.}{2007}]{daddi_multiwavelength_2007}
Daddi E.,  et~al., 2007, \mn@doi [ApJ] {10.1086/521818}, 670, 156

\bibitem[\protect\citeauthoryear{Daddi et~al.,}{Daddi et~al.}{2015}]{daddi_co_2015}
Daddi E.,  et~al., 2015, \mn@doi [A\&A] {10.1051/0004-6361/201425043}, 577, A46

\bibitem[\protect\citeauthoryear{Dale et~al.,}{Dale et~al.}{2023}]{dale_phangs-jwst_2023}
Dale D.~A.,  et~al., 2023, \mn@doi [ApJ] {10.3847/2041-8213/aca769}, 944, L23

\bibitem[\protect\citeauthoryear{Dekel et~al.,}{Dekel et~al.}{2009}]{dekel_cold_2009}
Dekel A.,  et~al., 2009, \mn@doi [Nature] {10.1038/nature07648}, 457, 451

\bibitem[\protect\citeauthoryear{Delvecchio et~al.,}{Delvecchio et~al.}{2021}]{delvecchio_infrared-radio_2021}
Delvecchio I.,  et~al., 2021, \mn@doi [A\&A] {10.1051/0004-6361/202039647}, 647, A123

\bibitem[\protect\citeauthoryear{Desai et~al.,}{Desai et~al.}{2007}]{desai_pah_2007}
Desai V.,  et~al., 2007, \mn@doi [ApJ] {10.1086/522104}, 669, 810

\bibitem[\protect\citeauthoryear{Ding et~al.,}{Ding et~al.}{2020a}]{ding_mass_2020}
Ding X.,  et~al., 2020a, \mn@doi [ApJ] {10.3847/1538-4357/ab5b90}, 888, 37

\bibitem[\protect\citeauthoryear{{Ding} et~al.,}{{Ding} et~al.}{2020b}]{2020ApJ...888...37D}
{Ding} X.,  et~al., 2020b, \mn@doi [\apj] {10.3847/1538-4357/ab5b90}, \href {https://ui.adsabs.harvard.edu/abs/2020ApJ...888...37D} {888, 37}

\bibitem[\protect\citeauthoryear{Dobbs, Theis, Pringle  \& Bate}{Dobbs et~al.}{2010}]{dobbs_simulations_2010}
Dobbs C.~L.,  Theis C.,  Pringle J.~E.,   Bate M.~R.,  2010, \mn@doi [MNRAS] {10.1111/j.1365-2966.2009.16161.x}, 403, 625

\bibitem[\protect\citeauthoryear{Donley, Rieke, Pérez-González  \& Barro}{Donley et~al.}{2008}]{donley_spitzers_2008}
Donley J.~L.,  Rieke G.~H.,  Pérez-González P.~G.,   Barro G.,  2008, \mn@doi [ApJ] {10.1086/591510}, 687, 111

\bibitem[\protect\citeauthoryear{Draine}{Draine}{2003}]{draine_interstellar_2003}
Draine B.~T.,  2003, \mn@doi [ARA\&A] {10.1146/annurev.astro.41.011802.094840}, 41, 241

\bibitem[\protect\citeauthoryear{Draine \& Li}{Draine \& Li}{2007}]{draine_infrared_2007}
Draine B.~T.,  Li A.,  2007, \mn@doi [ApJ] {10.1086/511055}, 657, 810

\bibitem[\protect\citeauthoryear{{Dunlop} et~al.,}{{Dunlop} et~al.}{2021}]{2021jwst.prop.1837D}
{Dunlop} J.~S.,  et~al., 2021, {PRIMER: Public Release IMaging for Extragalactic Research}, JWST Proposal. Cycle 1, ID. \#1837

\bibitem[\protect\citeauthoryear{Díaz-Santos et~al.,}{Díaz-Santos et~al.}{2011}]{diaz-santos_spatial_2011}
Díaz-Santos T.,  et~al., 2011, \mn@doi [The Astrophysical Journal] {10.1088/0004-637X/741/1/32}, 741, 32

\bibitem[\protect\citeauthoryear{Díaz-Santos et~al.,}{Díaz-Santos et~al.}{2017}]{diaz-santos_herschelpacs_2017}
Díaz-Santos T.,  et~al., 2017, \mn@doi [ApJ] {10.3847/1538-4357/aa81d7}, 846, 32

\bibitem[\protect\citeauthoryear{Egorov et~al.,}{Egorov et~al.}{2023}]{egorov_phangsjwst_2023}
Egorov O.~V.,  et~al., 2023, \mn@doi [ApJL] {10.3847/2041-8213/acac92}, 944, L16

\bibitem[\protect\citeauthoryear{Elbaz et~al.,}{Elbaz et~al.}{2011}]{elbaz_goodsherschel_2011}
Elbaz D.,  et~al., 2011, \mn@doi [A\&A] {10.1051/0004-6361/201117239}, 533, A119

\bibitem[\protect\citeauthoryear{Elbaz et~al.,}{Elbaz et~al.}{2018}]{elbaz_starbursts_2018}
Elbaz D.,  et~al., 2018, \mn@doi [A\&A] {10.1051/0004-6361/201732370}, 616, A110

\bibitem[\protect\citeauthoryear{{Elmegreen} \& {Elmegreen}}{{Elmegreen} \& {Elmegreen}}{2014}]{2014ApJ...781...11E}
{Elmegreen} D.~M.,  {Elmegreen} B.~G.,  2014, \mn@doi [\apj] {10.1088/0004-637X/781/1/11}, \href {https://ui.adsabs.harvard.edu/abs/2014ApJ...781...11E} {781, 11}

\bibitem[\protect\citeauthoryear{Elmegreen, Bournaud  \& Elmegreen}{Elmegreen et~al.}{2008}]{elmegreen_bulge_2008}
Elmegreen B.~G.,  Bournaud F.,   Elmegreen D.~M.,  2008, \mn@doi [ApJ] {10.1086/592190}, 688, 67

\bibitem[\protect\citeauthoryear{Faisst et~al.,}{Faisst et~al.}{2024}]{faisst_cosmos-web_2024}
Faisst A.~L.,  et~al., 2024, {COSMOS}-{Web}: {The} {Role} of {Galaxy} {Interactions} and {Disk} {Instabilities} in {Producing} {Starbursts} at z{\textless}4, \mn@doi{10.48550/arXiv.2405.09619}, \url {http://arxiv.org/abs/2405.09619}

\bibitem[\protect\citeauthoryear{Ferreira et~al.,}{Ferreira et~al.}{2023}]{ferreira_jwst_2023}
Ferreira L.,  et~al., 2023, \mn@doi [ApJ] {10.3847/1538-4357/acec76}, 955, 94

\bibitem[\protect\citeauthoryear{Förster~Schreiber \& Wuyts}{Förster~Schreiber \& Wuyts}{2020}]{forster_schreiber_star-forming_2020}
Förster~Schreiber N.~M.,  Wuyts S.,  2020, \mn@doi [ARA\&A] {10.1146/annurev-astro-032620-021910}, 58, 661

\bibitem[\protect\citeauthoryear{Genzel et~al.,}{Genzel et~al.}{2011}]{genzel_sins_2011}
Genzel R.,  et~al., 2011, \mn@doi [ApJ] {10.1088/0004-637X/733/2/101}, 733, 101

\bibitem[\protect\citeauthoryear{Gillman et~al.,}{Gillman et~al.}{2024}]{gillman_structure_2024}
Gillman S.,  et~al., 2024, \mn@doi [A\&A] {10.1051/0004-6361/202451006}, 691, A299

\bibitem[\protect\citeauthoryear{Greve et~al.,}{Greve et~al.}{2014}]{greve_star_2014}
Greve T.~R.,  et~al., 2014, \mn@doi [ApJ] {10.1088/0004-637X/794/2/142}, 794, 142

\bibitem[\protect\citeauthoryear{Grogin et~al.,}{Grogin et~al.}{2011}]{grogin_candels_2011}
Grogin N.~A.,  et~al., 2011, \mn@doi [ApJS] {10.1088/0067-0049/197/2/35}, 197, 35

\bibitem[\protect\citeauthoryear{Gullberg et~al.,}{Gullberg et~al.}{2019}]{gullberg_alma_2019}
Gullberg B.,  et~al., 2019, \mn@doi [MNRAS] {10.1093/mnras/stz2835}, 490, 4956

\bibitem[\protect\citeauthoryear{Hao, Kennicutt, Johnson, Calzetti, Dale  \& Moustakas}{Hao et~al.}{2011}]{hao_dust-corrected_2011}
Hao C.-N.,  Kennicutt R.~C.,  Johnson B.~D.,  Calzetti D.,  Dale D.~A.,   Moustakas J.,  2011, \mn@doi [ApJ] {10.1088/0004-637X/741/2/124}, 741, 124

\bibitem[\protect\citeauthoryear{Helou \& Bicay}{Helou \& Bicay}{1993}]{helou_physical_1993}
Helou G.,  Bicay M.~D.,  1993, \mn@doi [ApJ] {10.1086/173146}, 415, 93

\bibitem[\protect\citeauthoryear{Hodge \& da Cunha}{Hodge \& da~Cunha}{2020}]{hodge_high-redshift_2020}
Hodge J.~A.,  da Cunha E.,  2020, \mn@doi [Royal Society Open Science] {10.1098/rsos.200556}, 7, 200556

\bibitem[\protect\citeauthoryear{Hodge et~al.,}{Hodge et~al.}{2016}]{hodge_kiloparsec-scale_2016}
Hodge J.~A.,  et~al., 2016, \mn@doi [ApJ] {10.3847/1538-4357/833/1/103}, 833, 103

\bibitem[\protect\citeauthoryear{Hodge et~al.,}{Hodge et~al.}{2019}]{hodge_alma_2019}
Hodge J.~A.,  et~al., 2019, \mn@doi [ApJ] {10.3847/1538-4357/ab1846}, 876, 130

\bibitem[\protect\citeauthoryear{Ilbert et~al.,}{Ilbert et~al.}{2009}]{ilbert_cosmos_2009}
Ilbert O.,  et~al., 2009, \mn@doi [ApJ] {10.1088/0004-637X/690/2/1236}, 690, 1236

\bibitem[\protect\citeauthoryear{Inami et~al.,}{Inami et~al.}{2013}]{inami_mid-infrared_2013}
Inami H.,  et~al., 2013, \mn@doi [The Astrophysical Journal] {10.1088/0004-637X/777/2/156}, 777, 156

\bibitem[\protect\citeauthoryear{Jiménez-Andrade et~al.,}{Jiménez-Andrade et~al.}{2019}]{jimenez-andrade_radio_2019}
Jiménez-Andrade E.~F.,  et~al., 2019, \mn@doi [A\&A] {10.1051/0004-6361/201935178}, 625, A114

\bibitem[\protect\citeauthoryear{Kakiichi et~al.,}{Kakiichi et~al.}{2024}]{kakiichi_cosmos-3d_2024}
Kakiichi K.,  et~al., 2024, JWST Proposal. Cycle 3, p.~5893

\bibitem[\protect\citeauthoryear{Kalita, Silverman, Daddi, Bottrell, Ho, Ding  \& Yang}{Kalita et~al.}{2023}]{kalita_rest-frame_2023}
Kalita B.~S.,  Silverman J.~D.,  Daddi E.,  Bottrell C.,  Ho L.~C.,  Ding X.,   Yang L.,  2023, \mn@doi [ApJ] {10.3847/1538-4357/acfee4}, 960, 25

\bibitem[\protect\citeauthoryear{Kalita et~al.,}{Kalita et~al.}{2024}]{kalita_clumps_2024}
Kalita B.~S.,  et~al., 2024, \mn@doi [MNRAS] {10.1093/mnras/stae2781}

\bibitem[\protect\citeauthoryear{Kalita et~al.,}{Kalita et~al.}{2025a}]{kalita_multi-wavelength_2025}
Kalita B.~S.,  et~al., 2025a, A multi-wavelength investigation of spiral structures in \$z {\textgreater} 1\$ galaxies with {JWST}, \mn@doi{10.48550/arXiv.2501.03325}, \url {http://arxiv.org/abs/2501.03325}

\bibitem[\protect\citeauthoryear{Kalita, Silverman, Daddi, Mercier, Ho  \& Ding}{Kalita et~al.}{2025b}]{kalita_near-ir_2024}
Kalita B.~S.,  Silverman J.~D.,  Daddi E.,  Mercier W.,  Ho L.~C.,   Ding X.,  2025b, \mn@doi [MNRAS] {10.1093/mnras/staf031}, p. staf031

\bibitem[\protect\citeauthoryear{{Kashino} et~al.,}{{Kashino} et~al.}{2019}]{Kashino2019}
{Kashino} D.,  et~al., 2019, \mn@doi [\apjs] {10.3847/1538-4365/ab06c4}, \href {https://ui.adsabs.harvard.edu/abs/2019ApJS..241...10K} {241, 10}

\bibitem[\protect\citeauthoryear{Kennedy \& Eberhart}{Kennedy \& Eberhart}{1995}]{kennedy_particle_1995}
Kennedy J.,  Eberhart R.,  1995, in Proceedings of {ICNN}'95 - {International} {Conference} on {Neural} {Networks}. pp 1942--1948 vol.4, \mn@doi{10.1109/ICNN.1995.488968}, \url {https://ieeexplore.ieee.org/document/488968}

\bibitem[\protect\citeauthoryear{Kepley, Tsutsumi, Brogan, Indebetouw, Yoon, Mason  \& Donovan~Meyer}{Kepley et~al.}{2020}]{kepley_auto-multithresh_2020}
Kepley A.~A.,  Tsutsumi T.,  Brogan C.~L.,  Indebetouw R.,  Yoon I.,  Mason B.,   Donovan~Meyer J.,  2020, \mn@doi [PASP] {10.1088/1538-3873/ab5e14}, 132, 024505

\bibitem[\protect\citeauthoryear{Koekemoer et~al.,}{Koekemoer et~al.}{2007}]{koekemoer_cosmos_2007}
Koekemoer A.~M.,  et~al., 2007, \mn@doi [ApJS] {10.1086/520086}, 172, 196

\bibitem[\protect\citeauthoryear{Koekemoer et~al.,}{Koekemoer et~al.}{2011}]{koekemoer_candels_2011}
Koekemoer A.~M.,  et~al., 2011, \mn@doi [ApJS] {10.1088/0067-0049/197/2/36}, 197, 36

\bibitem[\protect\citeauthoryear{Kokorev et~al.,}{Kokorev et~al.}{2023}]{kokorev_dust_2023}
Kokorev V.,  et~al., 2023, \mn@doi [A\&A] {10.1051/0004-6361/202346937}, 677, A172

\bibitem[\protect\citeauthoryear{Kuhn, Guo, Martin, Bayless, Gates  \& Puleo}{Kuhn et~al.}{2024}]{kuhn_jwst_2024}
Kuhn V.,  Guo Y.,  Martin A.,  Bayless J.,  Gates E.,   Puleo A.~J.,  2024, \mn@doi [ApJL] {10.3847/2041-8213/ad43eb}, 968, L15

\bibitem[\protect\citeauthoryear{Lacki \& Thompson}{Lacki \& Thompson}{2010}]{lacki_physics_2010}
Lacki B.~C.,  Thompson T.~A.,  2010, \mn@doi [ApJ] {10.1088/0004-637X/717/1/196}, 717, 196

\bibitem[\protect\citeauthoryear{Lai, Smith, Baba, Spoon  \& Imanishi}{Lai et~al.}{2020}]{lai_all_2020}
Lai T. S.~Y.,  Smith J. D.~T.,  Baba S.,  Spoon H. W.~W.,   Imanishi M.,  2020, \mn@doi [ApJ] {10.3847/1538-4357/abc002}, 905, 55

\bibitem[\protect\citeauthoryear{Lai et~al.,}{Lai et~al.}{2022}]{lai_goals-jwst_2022}
Lai T. S.-Y.,  et~al., 2022, \mn@doi [ApJ] {10.3847/2041-8213/ac9ebf}, 941, L36

\bibitem[\protect\citeauthoryear{Lai et~al.,}{Lai et~al.}{2023}]{lai_goals-jwst_2023}
Lai T. S.-Y.,  et~al., 2023, \mn@doi [ApJ] {10.3847/2041-8213/ad0387}, 957, L26

\bibitem[\protect\citeauthoryear{Lai, Smith, Peeters, Spoon, Baba, Imanishi  \& Nakagawa}{Lai et~al.}{2024}]{lai_spectroscopic_2024}
Lai T. S.~Y.,  Smith J. D.~T.,  Peeters E.,  Spoon H. W.~W.,  Baba S.,  Imanishi M.,   Nakagawa T.,  2024, \mn@doi [ApJ] {10.3847/1538-4357/ad354b}, 967, 83

\bibitem[\protect\citeauthoryear{Lee et~al.,}{Lee et~al.}{2023}]{lee_phangs-jwst_2023}
Lee J.~C.,  et~al., 2023, \mn@doi [ApJ] {10.3847/2041-8213/acaaae}, 944, L17

\bibitem[\protect\citeauthoryear{Leger, D'Hendecourt, Boissel  \& Desert}{Leger et~al.}{1989}]{leger_photo-thermo-dissociation_1989}
Leger A.,  D'Hendecourt L.,  Boissel P.,   Desert F.~X.,  1989, A\&A, 213, 351

\bibitem[\protect\citeauthoryear{Li}{Li}{2020}]{li_spitzers_2020}
Li A.,  2020, \mn@doi [Nat Astron] {10.1038/s41550-020-1051-1}, 4, 339

\bibitem[\protect\citeauthoryear{Liu, Gao, Isaak, Daddi, Yang, Lu  \& Werf}{Liu et~al.}{2015}]{liu_high-j_2015}
Liu D.,  Gao Y.,  Isaak K.,  Daddi E.,  Yang C.,  Lu N.,   Werf P. v.~d.,  2015, \mn@doi [ApJL] {10.1088/2041-8205/810/2/L14}, 810, L14

\bibitem[\protect\citeauthoryear{Liu et~al.,}{Liu et~al.}{2019}]{liu_automated_2019}
Liu D.,  et~al., 2019, \mn@doi [ApJS] {10.3847/1538-4365/ab42da}, 244, 40

\bibitem[\protect\citeauthoryear{Liu et~al.,}{Liu et~al.}{2021}]{liu_co_2021}
Liu D.,  et~al., 2021, \mn@doi [ApJ] {10.3847/1538-4357/abd801}, 909, 56

\bibitem[\protect\citeauthoryear{Liu et~al.,}{Liu et~al.}{2023}]{liu_600_2023}
Liu D.,  et~al., 2023, \mn@doi [ApJ] {10.3847/1538-4357/aca46b}, 942, 98

\bibitem[\protect\citeauthoryear{Liu et~al.,}{Liu et~al.}{2024a}]{liu_detailed_2024}
Liu D.,  et~al., 2024a, \mn@doi [Nat Astron] {10.1038/s41550-024-02296-7}, 8, 1181

\bibitem[\protect\citeauthoryear{Liu et~al.,}{Liu et~al.}{2024b}]{liu_jwst_2024}
Liu Z.,  et~al., 2024b, \mn@doi [ApJ] {10.3847/1538-4357/ad4096}, 968, 15

\bibitem[\protect\citeauthoryear{Lyu et~al.,}{Lyu et~al.}{2024}]{lyu_primer_2024}
Lyu Y.,  et~al., 2024, \mn@doi [A\&A] {10.1051/0004-6361/202451067}

\bibitem[\protect\citeauthoryear{Madau \& Dickinson}{Madau \& Dickinson}{2014}]{madau_cosmic_2014}
Madau P.,  Dickinson M.,  2014, \mn@doi [ARA\&A] {10.1146/annurev-astro-081811-125615}, 52, 415

\bibitem[\protect\citeauthoryear{Magnelli et~al.,}{Magnelli et~al.}{2023}]{magnelli_ceers_2023}
Magnelli B.,  et~al., 2023, \mn@doi [A\&A] {10.1051/0004-6361/202347052}, 678, A83

\bibitem[\protect\citeauthoryear{Marchesi et~al.,}{Marchesi et~al.}{2016}]{marchesi_chandra_2016}
Marchesi S.,  et~al., 2016, \mn@doi [ApJ] {10.3847/0004-637X/817/1/34}, 817, 34

\bibitem[\protect\citeauthoryear{McKinney, Pope, Armus, Chary, Díaz-Santos, Dickinson  \& Kirkpatrick}{McKinney et~al.}{2020}]{mckinney_measuring_2020}
McKinney J.,  Pope A.,  Armus L.,  Chary R.-R.,  Díaz-Santos T.,  Dickinson M.~E.,   Kirkpatrick A.,  2020, \mn@doi [ApJ] {10.3847/1538-4357/ab77b9}, 892, 119

\bibitem[\protect\citeauthoryear{McKinney et~al.,}{McKinney et~al.}{2024}]{mckinney_scubadive_2024}
McKinney J.,  et~al., 2024, {SCUBADive} {I}: {JWST}+{ALMA} {Analysis} of 289 sub-millimeter galaxies in {COSMOS}-{Web}, \mn@doi{10.48550/arXiv.2408.08346}, \url {https://ui.adsabs.harvard.edu/abs/2024arXiv240808346M}

\bibitem[\protect\citeauthoryear{Murphy}{Murphy}{2013}]{murphy_role_2013}
Murphy E.~J.,  2013, \mn@doi [ApJ] {10.1088/0004-637X/777/1/58}, 777, 58

\bibitem[\protect\citeauthoryear{Murphy et~al.,}{Murphy et~al.}{2011}]{murphy_calibrating_2011}
Murphy E.~J.,  et~al., 2011, \mn@doi [ApJ] {10.1088/0004-637X/737/2/67}, 737, 67

\bibitem[\protect\citeauthoryear{Pedrini et~al.,}{Pedrini et~al.}{2024}]{pedrini_feedback_2024}
Pedrini A.,  et~al., 2024, \mn@doi [ApJ] {10.3847/1538-4357/ad534d}, 971, 32

\bibitem[\protect\citeauthoryear{{Perrin}, {Soummer}, {Elliott}, {Lallo}  \& {Sivaramakrishnan}}{{Perrin} et~al.}{2012}]{perrin_simulating_2012}
{Perrin} M.~D.,  {Soummer} R.,  {Elliott} E.~M.,  {Lallo} M.~D.,   {Sivaramakrishnan} A.,  2012, in {Clampin} M.~C.,  {Fazio} G.~G.,  {MacEwen} H.~A.,   {Oschmann} Jr. J.~M.,  eds,  Society of Photo-Optical Instrumentation Engineers (SPIE) Conference Series Vol. 8442, Space Telescopes and Instrumentation 2012: Optical, Infrared, and Millimeter Wave. p. 84423D, \mn@doi{10.1117/12.925230}

\bibitem[\protect\citeauthoryear{Polletta et~al.,}{Polletta et~al.}{2007}]{polletta_spectral_2007}
Polletta M.,  et~al., 2007, \mn@doi [ApJ] {10.1086/518113}, 663, 81

\bibitem[\protect\citeauthoryear{Polletta et~al.,}{Polletta et~al.}{2024}]{polletta_jwsts_2024}
Polletta M.,  et~al., 2024, \mn@doi [A\&A] {10.1051/0004-6361/202450671}, 690, A285

\bibitem[\protect\citeauthoryear{Pope et~al.,}{Pope et~al.}{2013}]{pope_probing_2013}
Pope A.,  et~al., 2013, \mn@doi [ApJ] {10.1088/0004-637X/772/2/92}, 772, 92

\bibitem[\protect\citeauthoryear{Puglisi et~al.,}{Puglisi et~al.}{2019}]{puglisi_main_2019}
Puglisi A.,  et~al., 2019, \mn@doi [ApJL] {10.3847/2041-8213/ab1f92}, 877, L23

\bibitem[\protect\citeauthoryear{Puglisi et~al.,}{Puglisi et~al.}{2021}]{puglisi_sub-millimetre_2021}
Puglisi A.,  et~al., 2021, \mn@doi [MNRAS] {10.1093/mnras/stab2914}, 508, 5217

\bibitem[\protect\citeauthoryear{Pérez-González et~al.,}{Pérez-González et~al.}{2024}]{perez-gonzalez_what_2024}
Pérez-González P.~G.,  et~al., 2024, \mn@doi [ApJ] {10.3847/1538-4357/ad38bb}, 968, 4

\bibitem[\protect\citeauthoryear{Rizzo et~al.,}{Rizzo et~al.}{2023}]{rizzo_alma-alpaka_2023}
Rizzo F.,  et~al., 2023, \mn@doi [A\&A] {10.1051/0004-6361/202346444}, 679, A129

\bibitem[\protect\citeauthoryear{Rodighiero et~al.,}{Rodighiero et~al.}{2011}]{rodighiero_lesser_2011}
Rodighiero G.,  et~al., 2011, \mn@doi [ApJL] {10.1088/2041-8205/739/2/L40}, 739, L40

\bibitem[\protect\citeauthoryear{Ronayne et~al.,}{Ronayne et~al.}{2024}]{ronayne_ceers_2024}
Ronayne K.,  et~al., 2024, \mn@doi [ApJ] {10.3847/1538-4357/ad5006}, 970, 61

\bibitem[\protect\citeauthoryear{Rujopakarn et~al.,}{Rujopakarn et~al.}{2023}]{rujopakarn_jwst_2023}
Rujopakarn W.,  et~al., 2023, \mn@doi [ApJL] {10.3847/2041-8213/accc82}, 948, L8

\bibitem[\protect\citeauthoryear{Shivaei \& Boogaard}{Shivaei \& Boogaard}{2024}]{shivaei_tight_2024}
Shivaei I.,  Boogaard L.~A.,  2024, \mn@doi [A\&A] {10.1051/0004-6361/202451826}, 691, L2

\bibitem[\protect\citeauthoryear{Shivaei et~al.,}{Shivaei et~al.}{2024}]{shivaei_new_2024}
Shivaei I.,  et~al., 2024, \mn@doi [A\&A] {10.1051/0004-6361/202449579}, 690, A89

\bibitem[\protect\citeauthoryear{Silverman et~al.,}{Silverman et~al.}{2015}]{silverman_higher_2015}
Silverman J.~D.,  et~al., 2015, \mn@doi [ApJL] {10.1088/2041-8205/812/2/L23}, 812, L23

\bibitem[\protect\citeauthoryear{Silverman et~al.,}{Silverman et~al.}{2018a}]{silverman_molecular_2018}
Silverman J.~D.,  et~al., 2018a, \mn@doi [ApJ] {10.3847/1538-4357/aae25e}, 867, 92

\bibitem[\protect\citeauthoryear{Silverman et~al.,}{Silverman et~al.}{2018b}]{silverman_concurrent_2018}
Silverman J.~D.,  et~al., 2018b, \mn@doi [ApJ] {10.3847/1538-4357/aae64b}, 868, 75

\bibitem[\protect\citeauthoryear{Smolčić et~al.,}{Smolčić et~al.}{2017}]{smolcic_vla-cosmos_2017}
Smolčić V.,  et~al., 2017, \mn@doi [A\&A] {10.1051/0004-6361/201628704}, 602, A1

\bibitem[\protect\citeauthoryear{Solomon \& Bout}{Solomon \& Bout}{2005}]{solomon_molecular_2005}
Solomon P.~M.,  Bout P. A.~V.,  2005, \mn@doi [ARA\&A] {10.1146/annurev.astro.43.051804.102221}, 43, 677

\bibitem[\protect\citeauthoryear{Speagle, Steinhardt, Capak  \& Silverman}{Speagle et~al.}{2014}]{speagle_highly_2014}
Speagle J.~S.,  Steinhardt C.~L.,  Capak P.~L.,   Silverman J.~D.,  2014, \mn@doi [ApJS] {10.1088/0067-0049/214/2/15}, 214, 15

\bibitem[\protect\citeauthoryear{Spergel}{Spergel}{2010}]{spergel_analytical_2010}
Spergel D.~N.,  2010, \mn@doi [ApJS] {10.1088/0067-0049/191/1/58}, 191, 58

\bibitem[\protect\citeauthoryear{Spilker et~al.,}{Spilker et~al.}{2023}]{spilker_spatial_2023}
Spilker J.~S.,  et~al., 2023, \mn@doi [Nature] {10.1038/s41586-023-05998-6}, 618, 708

\bibitem[\protect\citeauthoryear{Stierwalt et~al.,}{Stierwalt et~al.}{2013}]{stierwalt_mid-infrared_2013}
Stierwalt S.,  et~al., 2013, \mn@doi [The Astrophysical Journal Supplement Series] {10.1088/0067-0049/206/1/1}, 206, 1

\bibitem[\protect\citeauthoryear{Stierwalt et~al.,}{Stierwalt et~al.}{2014}]{stierwalt_mid-infrared_2014}
Stierwalt S.,  et~al., 2014, \mn@doi [The Astrophysical Journal] {10.1088/0004-637X/790/2/124}, 790, 124

\bibitem[\protect\citeauthoryear{Straatman et~al.,}{Straatman et~al.}{2016}]{straatman_fourstar_2016}
Straatman C. M.~S.,  et~al., 2016, \mn@doi [ApJ] {10.3847/0004-637X/830/1/51}, 830, 51

\bibitem[\protect\citeauthoryear{Tacconi et~al.,}{Tacconi et~al.}{2013}]{tacconi_phibss_2013}
Tacconi L.~J.,  et~al., 2013, \mn@doi [ApJ] {10.1088/0004-637X/768/1/74}, 768, 74

\bibitem[\protect\citeauthoryear{Tacconi, Genzel  \& Sternberg}{Tacconi et~al.}{2020}]{tacconi_evolution_2020}
Tacconi L.~J.,  Genzel R.,   Sternberg A.,  2020, \mn@doi [ARA\&A] {10.1146/annurev-astro-082812-141034}, 58, 157

\bibitem[\protect\citeauthoryear{Tan et~al.,}{Tan et~al.}{2024a}]{tan_situ_2024}
Tan Q.-H.,  et~al., 2024a, \mn@doi [Nature] {10.1038/s41586-024-08201-6}, 636, 69

\bibitem[\protect\citeauthoryear{Tan et~al.,}{Tan et~al.}{2024b}]{tan_fitting_2024}
Tan Q.-H.,  et~al., 2024b, \mn@doi [A\&A] {10.1051/0004-6361/202347255}, 684, A23

\bibitem[\protect\citeauthoryear{Tanaka et~al.,}{Tanaka et~al.}{2024}]{tanaka_m_rm_2024}
Tanaka T.~S.,  et~al., 2024, The \${M}\_\{{\textbackslash}rm {BH}\}-{M}\_*\$ relation up to \$z{\textbackslash}sim2\$ through decomposition of {COSMOS}-{Web} {NIRCam} images, \mn@doi{10.48550/arXiv.2401.13742}, \url {https://ui.adsabs.harvard.edu/abs/2024arXiv240113742T}

\bibitem[\protect\citeauthoryear{Teodoro \& Fraternali}{Teodoro \& Fraternali}{2015}]{teodoro_3dbarolo_2015}
Teodoro E. M.~D.,  Fraternali F.,  2015, \mn@doi [MNRAS] {10.1093/mnras/stv1213}, 451, 3021

\bibitem[\protect\citeauthoryear{Tielens}{Tielens}{2008}]{tielens_interstellar_2008}
Tielens A. G. G.~M.,  2008, \mn@doi [ARA\&A] {10.1146/annurev.astro.46.060407.145211}, 46, 289

\bibitem[\protect\citeauthoryear{Valentino et~al.,}{Valentino et~al.}{2020}]{valentino_co_2020}
Valentino F.,  et~al., 2020, \mn@doi [A\&A] {10.1051/0004-6361/202038322}, 641, A155

\bibitem[\protect\citeauthoryear{Zhang, Ho  \& Li}{Zhang et~al.}{2022}]{zhang_evidence_2022}
Zhang L.,  Ho L.~C.,   Li A.,  2022, \mn@doi [ApJ] {10.3847/1538-4357/ac930f}, 939, 22

\makeatother
\end{thebibliography}

%\begin{thebibliography}{99}
%\end{thebibliography}

%%%%%%%%%%%%%%%%%%%%%%%%%%%%%%%%%%%%%%%%%%%%%%%%%%

%%%%%%%%%%%%%%%%% APPENDICES %%%%%%%%%%%%%%%%%%%%%

\appendix

\section{An example of the foreground galaxy subtraction using S\'ersic modeling with Galight}

As depicted in Figure~\ref{fig:observations}, PACS-830 shows diverse substructures across different wavelengths, making it difficult to apply the same components of models in all the bands, while the foreground galaxy circled by the blue dash line in Figure~\ref{fig:observations} remains relatively consistent in terms of the shape from F435W to F150W. Therefore, we start from F814W, where the two galaxies have the best contrast, to model the foreground galaxy and sub-components of PACS-830 simultaneously. Afterward, we fix the position and S\'ersic shape of the foreground galaxy when fitting in the other bands. We set the fitting result of F606W as an example of the modeling and foreground subtraction.

\begin{figure*}
\centering
\includegraphics[width=14cm]{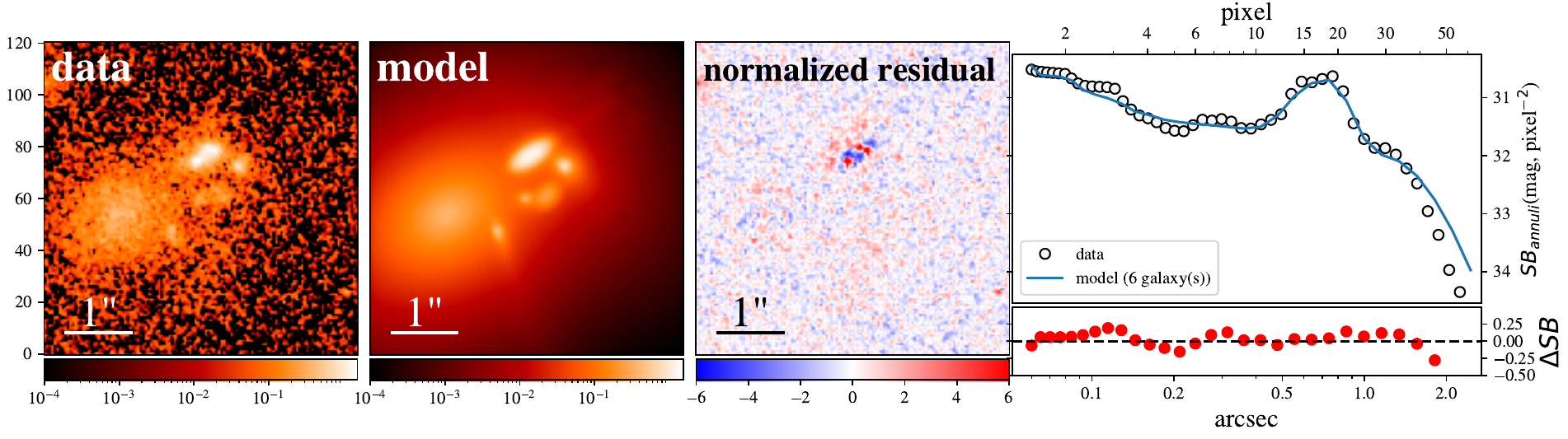}
\caption{S\'ersic model fit to the 2D emission in F606W using \texttt{Galight} with panels as follows from left to right: science image, model image, residual image (data--model/$\sigma$) and 1D surface brightness profile.}
\label{fig:img_model_fit}
\end{figure*}

\section{Moment 1 and 2 maps}\label{appendixB}
We generated the moment 1 (velocity field) and moment 2 (velocity dispersion) maps from the CO ($J$=5–4) data cube binned to 40$\,$km$\,$s$^{-1}$ channels. To ensure reliable measurements, we applied a mask that retains only those pixels with at least five consecutive channels having S/N above 5. For pixels adjacent to high-S/N regions, we relaxed the threshold to S/N~$>$~3.

The moment 1 map reveals a disk-like velocity field, consistent with rotation, and even shows hints of a spiral feature. The moment 2 map also aligns with expectations for a rotating disk, displaying elevated velocity dispersion in the central region. However, we note that the current spatial resolution and signal-to-noise ratio limit a more detailed interpretation. In particular, while the central enhancement in dispersion is qualitatively consistent with the presence of a bulge, it is also consistent with expectations for a disk without requiring a bulge component, unless a proper model comparison is made.

We also attempted to model the CO kinematics using the 3DBarolo code \citep{teodoro_3dbarolo_2015}. While the residual maps were generally acceptable, the fitted dispersion values carried large uncertainties and exceeded the measured values. This may reflect limitations in the modeling or a relatively low intrinsic velocity dispersion ($\lesssim$10 km/s). Furthermore, our spectral binning (40 km/s) is significantly larger than the expected dispersion, making the results unreliable. For these reasons, we decided not to include quantitative 3DBarolo results in either the main text or the appendix.

\begin{figure*}
\centering
\includegraphics[width=14cm]{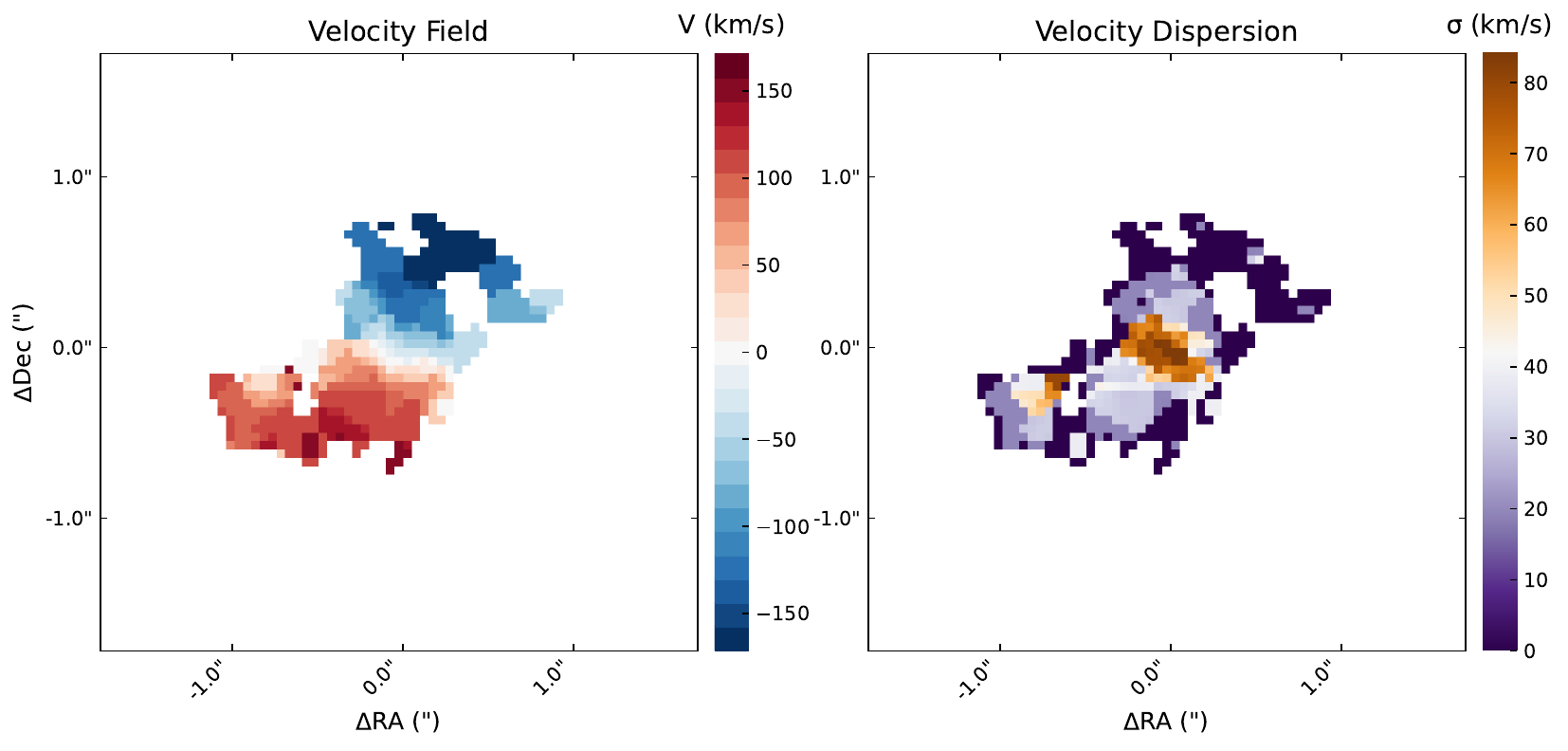}
\caption{Moment 1 and 2 map of CO ($J$=5--4).}
\label{fig:m1_2}
\end{figure*}

\section{VLA 3GHz observation}

\begin{figure}
\centering
\includegraphics[width=8cm]{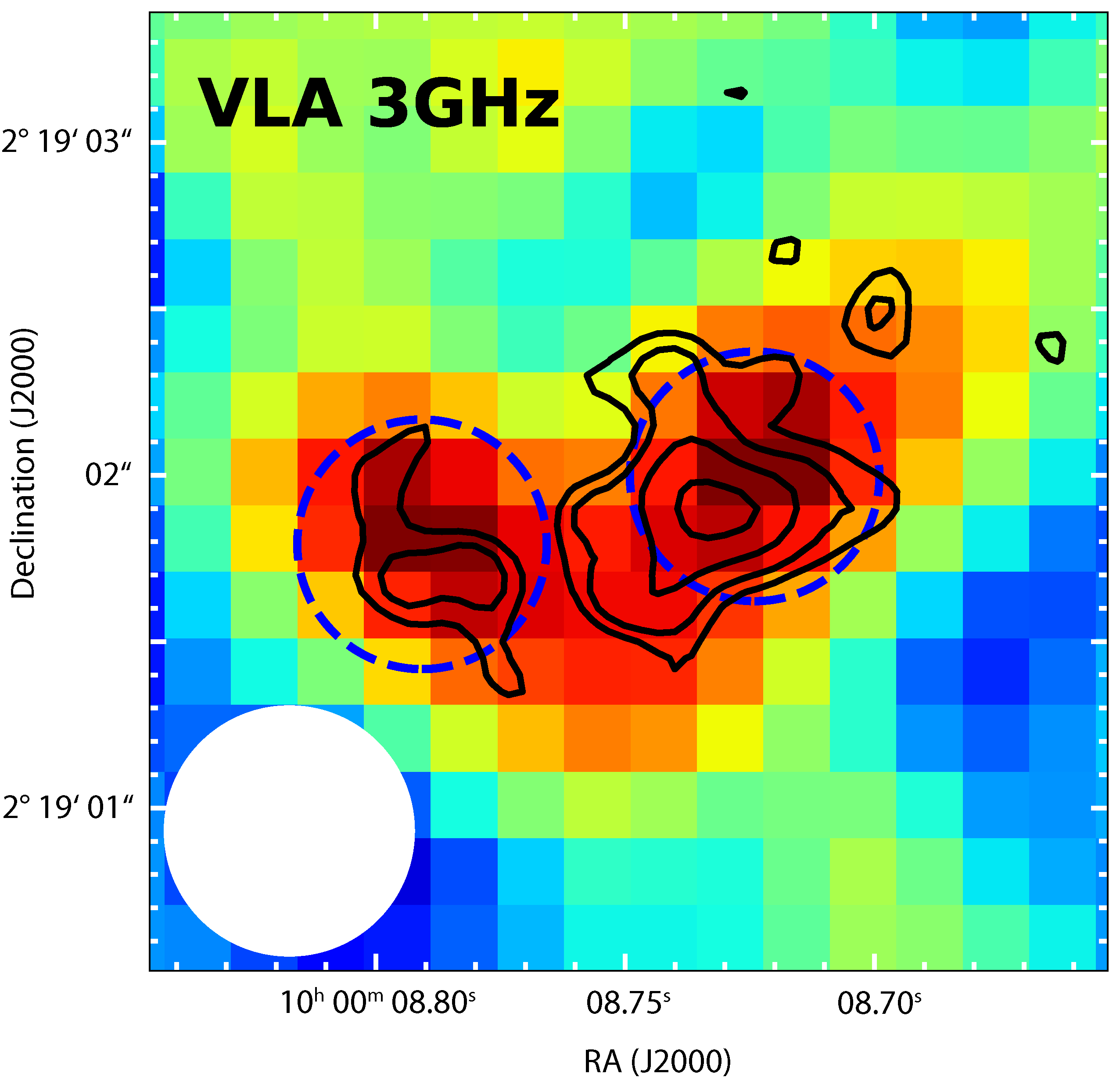}
\caption{VLA 3GHz observation. The blue circles represent the apertures of one beam size that we use to measure the flux ratio between the spiral arm and the core region. The black contours are generated from CO ($J$=5--4) emissions.}
\label{fig:alma_vla}
\end{figure}

We retrieve the VLA 3GHz observations from the VLA-COSMOS 3 GHz Large Project \citep{smolcic_vla-cosmos_2017} via the COSMOS cutout tool \footnote{\href{https://irsa.ipac.caltech.edu/data/COSMOS/index_cutouts.html}{COSMOS Cutouts}}. The beam size is 0$^{\prime \prime}$.75 $\times$ 0$^{\prime \prime}$.75. To our surprise, VLA captures two peaks equivalent in strength at the locations of the spiral arm and starbursting core, respectively. We measure the physical distance between the two peaks and the foreground galaxy to the galactic center of PACS-830 in F606W and confirm the eastern peak in VLA 3GHz is not from the foreground galaxy. As the peaks are slightly smaller than $5\sigma_\mathrm{rms}$ ($\sigma_\mathrm{rms}=2.39 \mu$Jy/beam), they are not included in the VLA-COSMOS catalog and the related paper on the radio sizes of SFGs \citep{smolcic_vla-cosmos_2017, jimenez-andrade_radio_2019}. Therefore, we will not overinterpret the data and only discuss the flux ratio ($\sim$1:1) in two different structures measured by two aperture photometries (r=0$^{\prime \prime}$.375, one beam size; locations shown in Figure~\ref{fig:alma_vla}). As a region with a high $\Sigma_\mathrm{SFR}$ and stellar mass density, the starburst core is expected to be of a lower IR-to-radio luminosity ratio $q_\mathrm{IR}$ \citep[for definition see eq.1 in][]{delvecchio_infrared-radio_2021} than the spiral arm region. Yet, the fact is vice versa. This could be explained by a flatter radio spectrum due to free-free absorption \citep[e.g.,][]{condon_compact_1991,murphy_role_2013} or a shorter cosmic ray scale heights \citep[e.g.,][]{helou_physical_1993,lacki_physics_2010}. Another explanation for the high radio emission in the spiral arm is that it could be a tidally disrupted galaxy hiding an AGN. Yet, this explanation can not address the enhanced radio emission between the two peaks.

A more certain investigation is beyond the scope of the data we currently have. We expect the next-generation radio telescope will solve the mystery of the missing radio flux in the starburst core of PACS-830.

%%%%%%%%%%%%%%%%%%%%%%%%%%%%%%%%%%%%%%%%%%%%%%%%%%

\newpage
\noindent
\textit{\footnotesize
$^{1}$Kavli Institute for the Physics and Mathematics of the Universe (Kavli IPMU, WPI), UTIAS, The University of Tokyo, Kashiwa, Chiba 277-8583, Japan\\
$^{2}$Department of Astronomy, School of Science, The University of Tokyo, 7-3-1 Hongo, Bunkyo, Tokyo 113-0033, Japan\\
$^{3}$Center for Data-Driven Discovery, Kavli IPMU (WPI), UTIAS, The University of Tokyo, Kashiwa, Chiba 277-8583, Japan\\
$^{4}$Université Paris-Saclay, Université Paris Cité, CEA, CNRS, AIM, F-91191 Gif-sur-Yvette, France\\
$^{5}$Center for Astrophysical Sciences, Department of Physics \& Astronomy, Johns Hopkins University, Baltimore, MD 21218, USA\\
$^{6}$Kavli Institute for Astronomy and Astrophysics, Peking University, Beijing 100871, P. R. China\\
$^{7}$School of Physics and Astronomy, University of Southampton, Highfield SO17 1BJ, UK\\
$^{8}$Department of Astronomy, School of Physics, Peking University, Beijing 100871, China\\
$^{9}$Istituto Nazionale di Astrofisica (INAF), Osservatorio Astronomico di Padova, Vicolo dell’Osservatorio 5, 35122, Padova, Italy\\
$^{10}$National Astronomical Observatory of Japan, 2-21-1 Osawa, Mitaka, Tokyo 181-8588, Japan\\
$^{11}$Cosmic Dawn Center (DAWN), Denmark\\
$^{12}$Niels Bohr Institute, University of Copenhagen, Jagtvej 128, DK-2200 Copenhagen N, Denmark\\
$^{13}$Laboratory for Multiwavelength Astrophysics, School of Physics and Astronomy, Rochester Institute of Technology, 84 Lomb Memorial Drive, Rochester, NY 14623, USA\\
$^{14}$Purple Mountain Observatory, Chinese Academy of Sciences, 10 Yuanhua Road, Nanjing 210023, China\\
$^{15}$Centro de Astrobiología (CAB), CSIC-INTA, Ctra. de Ajalvir km 4, Torrej\'on de Ardoz, 28850 Madrid, Spain\\
$^{16}$The University of Texas at Austin, 2515 Speedway Boulevard Stop C1400, Austin, TX 78712, USA\\
$^{17}$School of Physics and Technology, Wuhan University, Wuhan 430072, China\\
$^{18}$Caltech/IPAC, 1200 E. California Blvd., Pasadena, CA 91125, USA\\
$^{19}$Astrophysics Research, University of Hertfordshire, Hatfield, AL10 9AB, UK\\
$^{20}$Space Telescope Science Institute, 3700 San Martin Drive, Baltimore, MD 21218, USA\\
$^{21}$NASA Goddard Space Flight Center, Code 662, Greenbelt, MD 20771, USA\\
$^{22}$DTU Space, Technical University of Denmark, Elektrovej, Building 328, 2800 Kgs. Lyngby, Denmark\\
$^{23}$Department of Computer Science, Aalto University, P.O. Box 15400, FI-00076 Espoo, Finland\\
$^{24}$Department of Physics, University of, P.O. Box 64, FI-00014 Helsinki, Finland\\
$^{25}$Institut d'Astrophysique de Paris, UMR 7095, CNRS, and Sorbonne Université, 98 bis boulevard Arago, F-75014 Paris, France\\
$^{26}$Jet Propulsion Laboratory, California Institute of Technology, 4800 Oak Grove Drive, Pasadena, CA 91001, USA\\
$^{27}$Department of Astronomy and Astrophysics, University of California, Santa Cruz, 1156 High Street, Santa Cruz, CA 95064, USA\\
$^{28}$Department of Physics and Astronomy, Università degli Studi di Padova, Vicolo dell'Osservatorio 3, I-35122, Padova, Italy\\
$^{29}$Department of Physics, Faculty of Science, Chulalongkorn University, 254 Phayathai Road, Pathumwan, Bangkok 10330, Thailand\\
$^{30}$National Astronomical Research Institute of Thailand, 260 Moo 4, T. Donkaew, A. Maerim, Chiangmai 50180, Thailand\\
$^{31}$Astronomy Centre, University of Sussex, Falmer, Brighton BN1 9QH, UK\\
$^{32}$Department of Physics, University of Miami, Coral Gables, FL 33124, USA
}

% Don't change these lines
\bsp	% typesetting comment
\label{lastpage}
\end{document}